\documentclass[aps,pra,twocolumn,groupedaddress,showpacs,floatfix]{revtex4}
\usepackage{amsmath,amssymb,bm}
\usepackage{epsfig}
\usepackage{times,txfonts}

\renewcommand{\vec}[1]{\bm{\mathrm{#1}}}
\newcommand{\vecz}[1]{\bm{#1}}
\newcommand{\ket}[1]{| #1\rangle} 
\newcommand{\bra}[1]{\langle #1|}
\newcommand{\braket}[2]{\langle #1|#2\rangle}
\newcommand{\expect}[1]{\left\langle #1 \right\rangle}

\newcommand{\Hetrimer}{$^4$He$_3$}
\newcommand{\Hedimer}{$^4$He$_2$}
\newcommand{\Heatom}{$^4$He}

\begin{document}

\title{Matter diffraction at oblique incidence: Higher resolution and
  the {\Hetrimer} Efimov state}

\author{Gerhard C.~Hegerfeldt}\author{Martin Stoll}
\affiliation{Institut f{\"u}r Theoretische Physik, Universit{\"a}t
  G{\"o}ttingen, Friedrich-Hund-Platz 1, 37077 G{\"o}ttingen,
  Germany}

\date{\today}

\begin{abstract}
  We study the diffraction of atoms and weakly-bound three-atomic
  molecules from a transmission grating at non-normal incidence. Due
  to the thickness of the grating bars the slits are partially
  shadowed. Therefore, the projected slit width decreases more
  strongly with the angle of incidence than the projected period,
  increasing, in principle, the experimental resolution.  The
  shadowing, however, requires a revision of the theory of atom
  diffraction. We derive an expression in the style of the Kirchhoff
  integral of optics and show that the diffraction pattern exhibits a
  characteristic asymmetry which must be accounted for when comparing
  with experimental data. We then analyze the diffraction of weakly
  bound trimers and show that their finite size manifests itself in a
  further reduction of the slit width by $(3/4)\expect{r}$ where
  $\expect{r}$ is the average bond length.  The improved resolution at
  non-normal incidence may in particular allow to discern, by means of
  their bond lengths, between the small ground state of the helium
  trimer ($\expect{r}\approx1\,$nm, Barletta and Kievsky, Phys.~Rev.~A
  \textbf{64}, 042514 (2001)) and its predicted Efimov-type excited
  state ($\expect{r}\approx8\,$nm, ibid.), and in this way to
  experimentally prove the existence of this long-sought Efimov state.
\end{abstract}


\maketitle

\section{Introduction\label{sec:introduction}}

The combination of two unique features makes matter-wave diffraction
of noble gas trimers an outstanding enterprise.  Firstly, diffraction
presently is the only experimental technique which allows to detect
such very weakly bound clusters and to determine their properties
\cite{LGG_JCP104, ST_SCIENCE266, GSTHKS_PRL85}. Secondly, the helium
trimer {\Hetrimer} is the only molecule predicted to possess an
Efimov-type bound state \cite{LDD_PRL38,ELG_PRA54,BK_PRA64} under
normal conditions \cite{footnote-ultracold}.

Only recently did diffraction of atomic and molecular beams evolve
towards a precise experimental technique.  Early pioneering
experiments had proved diffraction for a sodium beam through a grating
initially fabricated for x-rays \cite{KSSP_PRL61}, as well as for
meta-stable helium \cite{CM_PRL66} and for neon \cite{SST_PRA46}
through micrometer double-slits. Later, transmission gratings with a
period of $d=200$\,nm had brought an improvement, but finally the
production of reliable nano-scale transmission gratings with a period
of only $d=100$\,nm \cite{SSCS_JVSTB13} paved the way for quantitative
matter diffraction experiments with an unparalleled spatial coherence
across up to a hundred slits for a helium atom beam
\cite{GSTHK_PRL83}.  Unprecedented, matter diffraction also allows to
investigate the coherence properties of very heavy molecules with many
internal degrees of freedom such as fullerenes
\cite{ANVKZA_N401,HUBHAZ_PRL90}.

Classical wave optics can merely serve as an approximation to the
underlying physical scattering process of atom diffraction.  The
hierarchy of the diffraction peak intensities in a diffraction
experiment with neutral atoms or molecules is significantly affected
by the weak van der Waals surface force, which acts on atoms in the
vicinity of the material grating. This was included in a quantitative
theory in Refs.~\cite{HK_PRA57, HK_PRA61} which allowed, in comparison
with experimental data, to characterize this surface force for the
noble gases helium, neon, argon, and krypton, and the covalently bound
D$_2$ molecule \cite{GSTHK_PRL83}, as well as for meta-stable helium
and neon \cite{BFGTHKSW_EPL59}.

Moreover, there is no analog in wave optics for the diffraction of
weakly bound small noble gas van der Waals clusters such as the helium
dimer {\Hedimer} and trimer {\Hetrimer}.  Experimental evidence for
these delicate molecules was for the first time unambiguously provided
through the mass-selective property of grating diffraction
\cite{ST_SCIENCE266}.  Moreover, the comparatively large interatomic
distance of $\expect{r}=5.2$\,nm in {\Hedimer}, implied by the small
binding energy $E_\text{b}=-1.1$\,mK \cite{GSTHKS_PRL85}, was shown to
manifest itself as an apparent narrowing of the grating slits by
$\frac12\expect{r}$.  It was this size effect which rendered possible
the determination of $\expect{r}$ from experimental diffraction data
\cite{GSTHKS_PRL85}.  This allowed for comparison with modern
realistic helium-helium potentials of
Refs.~\cite{JA_JCP103,G_MOLPHYS99}.

Numerical studies of the helium trimer relying on these helium-helium
potentials \cite{LDD_PRL38,HL_JCP68,NL_PRA28,FJ_PRL71,NFJ_JPB31,
  MSK_CPL275,KMS_JPB31,RY_CPL328,MSSK_JPB31,ELG_PRA54,B_JCP110,BK_PRA64},
quantum Monte Carlo simulations \cite{L_JCP106}, as well as quantum
chemical \emph{ab initio} calculations \cite{RA_JCP102} have long
predicted two bound states for {\Hetrimer}: a ground state at
$-126$\,mK and a shallow excited state at $-2.3$\,mK \cite{BK_PRA64}.
Moreover, the excited state is believed to be an Efimov-type state.
Originally in the context of nuclear physics, Efimov \cite{E_PL33B}
had shown that if the scattering length of a pair potential exceeds
the effective range of the potential by far then a universal series of
bound states exists in the three-body system near the dissociation
threshold.  Examples for such Efimov states, however, have been
searched for in vain in three-nucleon systems, leaving the
three-atomic helium molecule presently as the only candidate.

The {\Hetrimer} excited state cannot experimentally be distinguished
from the ground state by its mass. However, due to the large
difference in binding energy, both predicted states have markedly
different interatomic distances: $\expect{r}=0.96$\,nm in the ground
state and $\expect{r}=7.97$\,nm in the excited state \cite{BK_PRA64}.
Therefore, the size effect, which had previously played the essential
role in the dimer diffraction experiment and which we show to be
$\frac34\expect{r}$ for a trimer, is expected to render the two states
distinguishable.  Bruch \emph{et al} have analyzed the mole fraction
of small helium clusters (including atoms) in a nozzle beam
diffraction setup and showed that up to 7\% can be trimers
\cite{BST_JCP117}. This should be an ample amount for a quantitative
analysis. The population ratio of ground state vs.~excited state
trimers in the beam is, however, not known. It is therefore essential
to provide sufficient experimental resolution for the small ground
state in order to evaluate diffraction data from a mixed beam. A
limitation in the resolution is posed by the period of the grating,
$d=100\,$nm, and the slit width, typically $s_0=60$\,nm, which are
both large compared to the ground state size.  Transmission gratings
with smaller periods and slit widths are, however, presently not
available.

To address this issue we consider diffraction from a custom
transmission grating at oblique (non-normal) incidence, i.e.~the
grating is rotated by an angle $\theta'$ about an axis parallel to its
bars. The grating typically consists of a $t\approx 120$\,nm thick
layer of silicon nitride into which the slits have been etched in a
lithographic production process \cite{SSCS_JVSTB13}. As the grating is
rotated the upstream edges of its bars cut into the incident beam,
partially shadowing the slits.  By this effect the projected slit
$s_\perp$ width (perpendicular to the incident beam) decreases more
strongly with $\theta'$ than the projected period $d_\perp$.  Since the
diffraction pattern is roughly governed by the squared slit function
\cite{bornwolf}
\begin{displaymath}
  \left[\frac{\sin(n\pi s_\perp/d_\perp)}
    {(n\pi s_\perp/d_\perp)}\right]^2
\end{displaymath}
where $n$ denotes the diffraction order, one expects that at
non-normal incidence more atoms or molecules are diffracted into
higher diffraction orders. For example, at $t=120\,$nm and an angle of
incidence (rotation angle) of $\theta'=21^\circ$ the ratio
$d_\perp/s_\perp$ is twice as large as $d/s_0$ (normal incidence)
while the diffraction angles, to good approximation, only increase by
the ratio $d/d_\perp$.  This effect increases the experimental
resolution. Due to the shadowing, however, the fundamental results
obtained earlier for atom diffraction at normal incidence cannot be
carried over unchanged.

While we illustrate our results using the experimentally most
interesting case of helium trimer diffraction, the general findings of
this work equally apply to other weakly-bound trimers, possibly
consisting of non-identical atoms.  The article is structured as
follows.  In Section \ref{sec:generaltheory} we derive, from quantum
mechanical scattering theory, the transition amplitude for an atom
diffracted from a bar of finite thickness. In Section
\ref{sec:atomtiltedgrating} we construct a periodic transmission
grating from many bars and introduce the notion of a slit at
non-normal incidence. We show that if the slits are not aligned with
the direction of periodicity a characteristic asymmetry of the
diffraction pattern arises which went unnoticed in a previous
experiment \cite{GSTMSS_PRA61}. The asymmetry is relevant for the
precise evaluation of experimental data. In Section
\ref{sec:trimerscatteringtheory} we outline the general scattering
theory approach to trimer diffraction and work it out in Section
\ref{sec:trimertiltedgrating} for non-normal trimer diffraction from a
grating. In Section \ref{sec:trimersize} we provide the link between
diffraction data and the trimer bond length and discuss aspects of a
helium trimer diffraction experiment.

\section{Atom diffraction from a deep bar\label{sec:generaltheory}}

In a typical beam diffraction experiment with an average beam velocity
of the order of $v=500$\,m/s the kinetic energy per atom is a few tens
of meV, much less than electronic excitation energies of the atom.
Therefore, we treat atoms as point particles and neglect their
electronic degrees of freedom throughout this article.  The de Broglie
wavelength $\lambda_\text{dB}$ associated with the atomic motion is
typically of the order of $0.1\,$nm whereas a typical length scale of
the scattering object is $d=100$\,nm. We shall, in the following,
refer to this relation as the \emph{diffraction condition}:
\begin{equation}
  \label{eq:diffractioncondition}
  \lambda_\text{dB}\ll d\ .
\end{equation}

The free Hamilton operator for an atom of mass $m$ is $H_0=\hat{\vec
  p}^2/2m$. The interaction between the diffracting object and the
atom will be described by a Lennard-Jones \cite{L_TFS28} type surface
potential $W(\vec x)$ where $\vec x$ is the position of the atom. This
interaction exhibits a strongly repulsive core at a distance $l$ from
the diffracting object of the order of the atomic diameter, and it
passes into a weak attractive $-C_3/l^3$ van der Waals potential at
$l\gtrsim1$\,nm \cite{RK_PR179,GSTHK_PRL83}. Due to the low kinetic
energy of the atoms in the beam it is sufficient for the purposes of
this work to model the repulsive part of the interaction by a hard
core. The attractive part will be omitted for the moment and will be
be included later in Sec.~\ref{sec:atomtiltedgrating}.

Generally, the scattering state $\ket{\vec p',+}$ for an atom
with incident momentum $\vec p'$ and positive energy $E'=|\vec
p'|^2/2m$ satisfies the Lippmann-Schwinger equation \cite{newton}
\begin{equation}
  \label{eq:atomLSG}
  \ket{\vec p',+}=\ket{\vec p'}+G_0(E'+ i 0)W\ket{\vec p',+}
\end{equation}
where $G_0(z)=\left[z-H_0\right]^{-1}$ is the free Green's function,
or resolvent.  Denoting the atom transition amplitude associated with
the potential $W$ by
\begin{equation}
  \label{eq:atomstreuamplitude}
  t^\text{at}(\vec p;\vec p')=\bra{\vec p}W\ket{\vec p',+},
\end{equation}
where $\vec p=\vec p'+\Delta\vec p$ is the outgoing momentum and
$E=|\vec p|^2/2m$, the $S$ matrix element has the usual decomposition
\cite{newton}
\begin{equation}
  \label{eq:Smatrix}
  \bra{\vec p} S \ket{\vec p'}
  =\delta^{(3)}(\vec p-\vec p') 
  - 2\pi i \delta(E-E')\ t^\text{at}(\vec p;\vec p').
\end{equation}

In many applications the diffraction object may be effectively
regarded as translationally invariant along one direction whence the
diffraction process can be treated in two space dimensions. This is
the case, for instance, for diffraction from a slit if the vertical
spread of the focussed atom beam is much less than the physical height
of the slit. In this article we shall always assume the scattering
object to be translationally invariant along the $x_3$-axis. To adapt
the notation, we denote the two-dimensional projections into the
$(x_1,x_2)$ plane of all three-dimensional vectors, such as $\vec p$,
by their corresponding italic letters, such as $\vecz p$. As
scattering does not occur along the $x_3$-axis a delta-function
$\delta(\Delta p_3)$ expressing momentum conservation can be extracted
from Eq.~(\ref{eq:atomstreuamplitude}), leaving a two-dimensional
transition amplitude $t^{\text{at}{(2)}}(\vecz p;\vecz p')$ which
satisfies
\begin{equation}
  \label{eq:atomstreuamplitude3D_2D}
  t^\text{at}(\vec p;\vec p')=\delta(\Delta p_3)\ 
  t^{\text{at}{(2)}}(\vecz p;\vecz p').
\end{equation}
In order to derive an expression for $t^{\text{at}{(2)}}(\vecz p;\vecz
p')$ we project Eq.~(\ref{eq:atomLSG}) into configuration space.  The
full wave function $\psi(\vecz k',\vecz x)=2\pi\hbar\braket{\vecz
  x}{\vecz p',+}$, where $\vecz k'=\vecz p'/\hbar$, is a sum of the
incident part $\psi_\text{inc}(\vecz k',\vecz
x)=2\pi\hbar\braket{\vecz x}{\vecz p'}$ and the scattered part
\begin{equation}
  \label{eq:psiscatt}
  \psi_\text{scatt}(\vecz k',\vecz x)=2\pi\hbar \bra{\vecz x}
  G^{(2)}_0\left(E'^{(2)}+ i 0\right)W\ket{\vecz p',+}.
\end{equation}
Here, $E'^{(2)}=|\vecz p'|^2/2m$, and the two-dimensional Green's
function, or resolvent, is $G^{(2)}_0(z)=[z-\hat{\vecz p}^2/2m]^{-1}$. 

\begin{figure}[htbp]
  \centering
  \epsfig{file=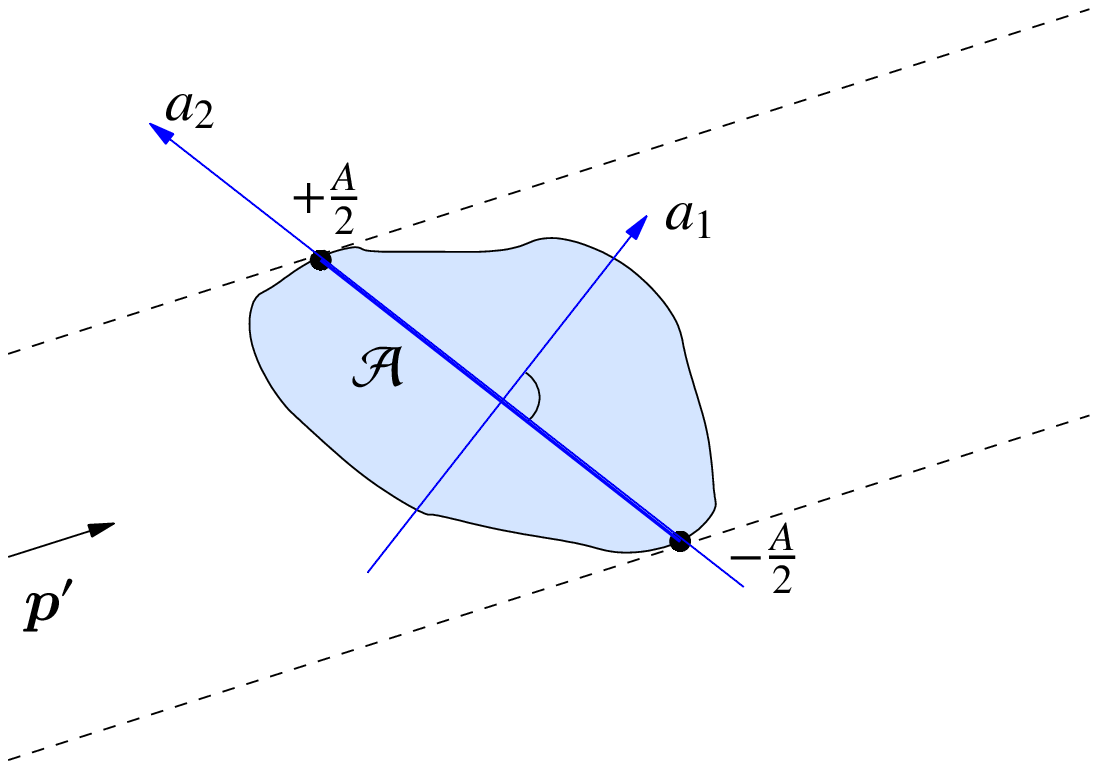,width=\columnwidth}
  \caption{Two-dimensional geometry of atom diffraction from a 
    diffraction object (bar) of finite thickness. The incident
    momentum of the atom is denoted by $\vecz p'$.  The straight
    shadow line $\mathcal A$, which plays the role of the silhouette,
    divides the bar into an ``illuminated'' part and a ``shadowed''
    part.  Also drawn is the adapted coordinate system $(a_1, a_2)$
    with the $a_2$-axis centered along the shadow line.}
  \label{fig:atom_scattering}
\end{figure}

If the scale of an object is large compared to the wavelength of
visible light it is well known that the diffraction about the forward
direction depends only on its (two-dimensional) silhouette as seen
from the direction of the illuminating light, e.g.~as for a disk and a
ball.  In two dimensions the silhouette of a diffraction object is
simply a straight line, here called shadow line (line $\mathcal A$ in
Fig.~\ref{fig:atom_scattering}) and, as in optics, due to the
diffraction condition (\ref{eq:diffractioncondition}) the scattered
part $\psi_\text{scatt}$ of the wave function can be approximated at
small scattering angles about the forward direction
\cite{morsefeshbach-diffr}. Neglecting the attractive part of
the potential $W(\vec x)$ the repulsive hard core imposes Dirichlet
boundary conditions on the circumference of the diffracting object.
Denoting the Green's function in configuration space by
\begin{equation}
  \label{eq:atomResolvOrt}
  G^{(2)}_0(|\vecz k'|;\vecz x,\vecz x')=-\frac{\hbar^2}{2m}
  \bra{\vecz x}G^{(2)}_0\left(E'^{(2)}+ i 0\right)\ket{\vecz x'}
\end{equation}
and using the Green theorem one finds, after some algebra,
\begin{eqnarray}
  \nonumber
  \psi_\text{scatt}(\vecz k',\vecz x)&\simeq&
  \int_{\mathcal{A}} d a_2
  \left[
    G^{(2)}_0(|\vecz k'|;\vecz x,\vecz a){\frac{\partial}{\partial a_1}}
    \psi_\text{inc}(\vecz k',\vecz a)
  \right.\\
  \label{eq:psidiffr_0}
  & &\ \left.
    -\psi_\text{inc}(\vecz k',\vecz a){\frac{\partial}{\partial a_1}}
    G^{(2)}_0(|\vecz k'|;\vecz x,\vecz a)
  \right]_{a_1=0}.
\end{eqnarray}
Here, $ d a_2$ is the infinitesimal line element along $\mathcal{A}$,
and $\partial/\partial a_1$ denotes the normal derivative
(cf.~Fig.~\ref{fig:atom_scattering}). The Green's function
(\ref{eq:atomResolvOrt}) can be expressed in terms of a Hankel
function \cite[chap.~3.10]{coltonkress}:
\begin{equation}
  \label{eq:atomResolvHankel}
  G^{(2)}_0(|\vecz k'|;\vecz x,\vecz x')=
  \frac{i}4 H^{(1)}_0\left(|\vecz k'||\vecz x-\vecz x'|\right).
\end{equation}
Using the asymptotic expansion \cite[chap.~9.2]{abramowitzsegun}
\begin{equation}
  \label{eq:hankelasymp}
  H^{(1)}_0\left(|\vecz k'||\vecz x-\vecz x'|\right)
  \stackrel{|\vecz x|\rightarrow \infty}{\sim}
  \sqrt{\frac{2}{\pi |\vecz k'|}}\ 
  e^{-i|\vecz k'| \vecz x'\cdot\vecz x/|\vecz x|}\
  \frac{e^{i(|\vecz k'||\vecz x|-\pi/4)}}{\sqrt{|\vecz x|}}
\end{equation}
the far field (Fraunhofer) limit of $\psi_\text{scatt}(\vecz
k',\vecz x)$ can readily be calculated. Inserting the expansion into
Eq.~(\ref{eq:psidiffr_0}) shows that the vector $\vecz k=|\vecz
k'|\vecz x/|\vecz x|$ in the first exponential should be identified with
the outgoing wave vector, and $\vecz p=\hbar \vecz k$ with the outgoing
momentum.  Comparing this expression with the far field limit of
Eq.~(\ref{eq:psiscatt}) one arrives at the two-dimensional transition
amplitude
\begin{subequations}
\begin{align}
  \label{eq:atom_transampl_bar}
  t^{\text{at}{(2)}}(\vecz p;\vecz p')&=
  -\frac{i}2\frac{p_{a_1}+p'_{a_1}}{(2\pi)^2m\hbar}
  \int_{-A/2}^{A/2} d a_2
  \ e^{- i\Delta p_{a_2} a_2/\hbar}\\
  &=-\frac{i}2\frac{p_{a_1}+p'_{a_1}}{(2\pi)^2m\hbar}
  \frac{\sin(\Delta p_{a_2}A/2\hbar)}{\Delta p_{a_2}/2\hbar},
\end{align}
\end{subequations}
where $A$ denotes the length of the shadow line $\mathcal{A}$.
Furthermore, $p'_{a_1}$ is the momentum component of $\vecz p'$ normal
to the shadow line, and $p'_{a_2}$ is the parallel component
(cf.~Fig.~\ref{fig:atom_scattering}). In accordance with the Babinet
principle, the transition amplitude shows the characteristic behavior
of an optical slit function.  The Babinet principle of wave optics
states that two complementary objects, such as a slit and a bar of the
same width, cause the same diffraction pattern outside the direction
of illumination (forward scattering) \cite{bornwolf}.

\section{Atom diffraction from a deep
  grating \label{sec:atomtiltedgrating}}

\subsection{The atom slit function of the deep grating}

\begin{figure}[htbp]
  \centering
  \epsfig{file=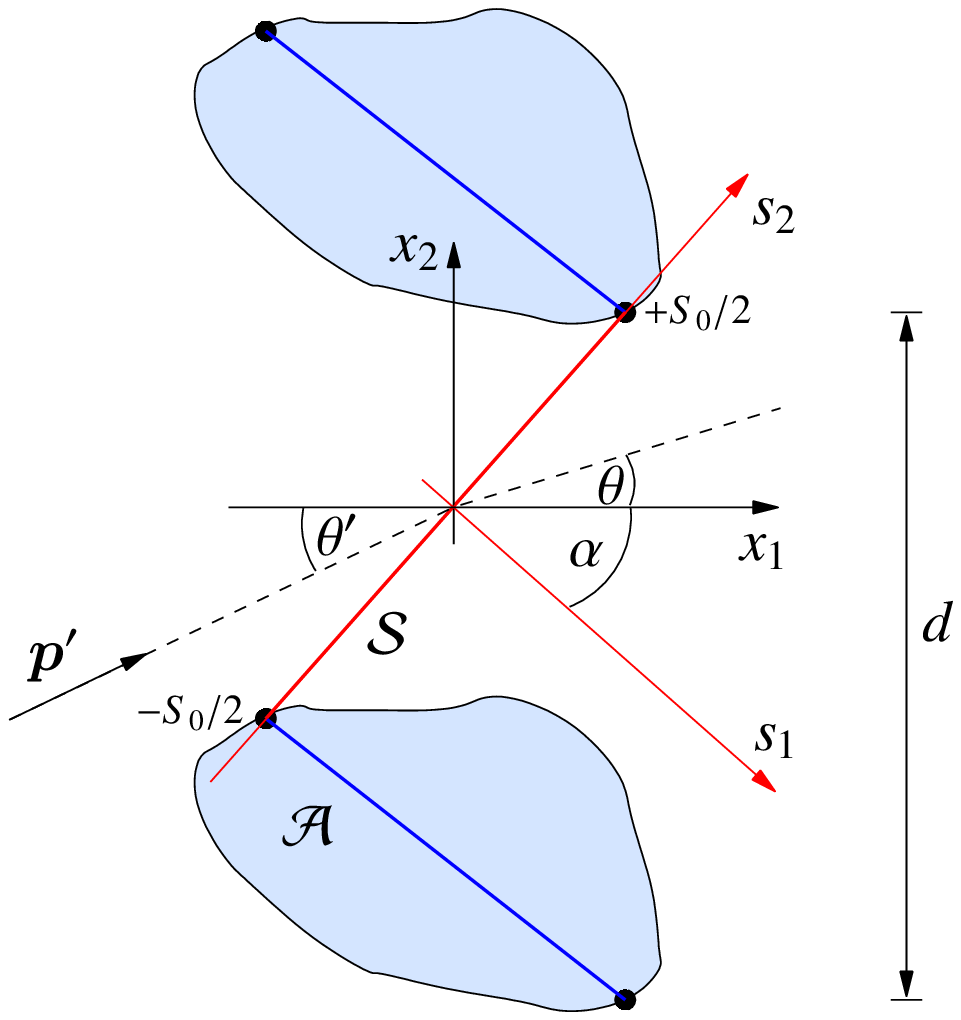,width=\columnwidth}
  \caption{Geometry of atom diffraction from a grating
    of identical deep bars with period $d$ along the $x_2$ direction.
    The slit line $\mathcal S$, which connects the shadow lines
    $\mathcal A$ of adjacent bars, provides a generalization of the
    notion of a slit at normal incidence. It lends itself to the
    application of the Huygens principle. Also drawn is the adapted
    coordinate system $(s_1,s_2)$ with the $s_2$-axis centered along
    the slit line and the angle $\alpha$ by which the coordinate
    systems $(x_1,x_2)$ and $(s_1,s_2)$ are rotated with respect to
    each other. }
  \label{fig:atom_scattering_grating}
\end{figure}

In diffraction experiments one often employs transmission gratings to
enhance the measurable diffraction peak intensities by a factor of
$N^2$, where $N$ is the number of coherently illuminated bars. One
grating bar is simply a special case of the general scattering object
considered in the previous section. In the following we arrange $N$
identical bars to create a regularly spaced periodic transmission
grating. While the simplest and most familiar situation in which the
bars are aligned along their common shadow line ($a_2$-axis,
cf.~Fig.~\ref{fig:atom_scattering}) is treated in virtually every
textbook on optics (e.g.~Ref.~\cite{bornwolf}), we are not aware of a
more general treatment where the individual shadow lines are not
parallel to the alignment axis. This situation arises naturally,
however, for diffraction from a grating with bars of finite thickness
at non-zero angle of incidence
(cf.~Fig.~\ref{fig:atom_scattering_grating}). We note that apart from
the period an additional length scale of the grating is given by the
distance between the bars. Depending on the angle of incidence the
(projected) distance can become small or even zero. The diffraction
condition (\ref{eq:diffractioncondition}), which must hold for all
length scales, therefore also imposes a limit on the maximal angle of
incidence.

Under the diffraction condition the transition amplitude of a grating
of $N$ bars with period $d$ along the $x_2$-axis can be written as the
coherent sum of the spatially translated amplitudes of each bar:
\begin{subequations}
  \label{eq:atom_transampl_gra}
  \begin{align}
    \label{eq:atom_transampl_gra_a}
    t^{\text{at}{(2)}}_\text{gra}(\vecz p;\vecz p')&=
    \sum_{n=0}^{N-1} e^{-i(n-\frac12(N-1))\Delta p_2 d/\hbar} \
    t^{\text{at}{(2)}}(\vecz p;\vecz p')\\
    &=H_N(\Delta p_2)\ t^{\text{at}{(2)}}(\vecz p;\vecz p').
  \end{align}
\end{subequations}
In the second line, the sum has been carried out and replaced by the
grating function \cite{bornwolf}
\begin{equation}
  \label{eq:gitterfunktion}
  H_N(\Delta p_2)=\frac{\sin(\Delta p_2 d N/2\hbar)}
  {\sin(\Delta p_2 d /2\hbar)}
\end{equation}
whose argument is the momentum transfer along the direction of
periodicity $x_2$. Eq.~(\ref{eq:atom_transampl_gra}) yields, in
principle, a satisfactory description of the diffraction problem in
terms of atom scattering from a bar. The literature on optics,
however, commonly adapts a complementary point of view by focusing on
the apertures (slits) between the bars rather than on the aperture
stops (bars) themselves.  This is expressed, for example, by the
Kirchhoff integral of optics \cite{bornwolf}.  In recent work this
viewpoint has also proven very useful in the field of atom and
molecule diffraction from a transmission grating: small quantum
mechanical effects such as the van der Waals interaction
\cite{GSTHK_PRL83,BFGTHKSW_EPL59} between the atoms in the beam and
the grating as well as the finite size of the helium dimer
\cite{GSTHKS_PRL85} manifest themselves as an apparent reduction of
the slit width of the grating. In these articles the respective
quantities could be determined quite precisely by comparison of the
reduced slit width with the true geometrical slit width. Unlike for
grating diffraction at normal incidence, however, it is initially not
evident how to define a slit in the case of non-normal incidence: the
correct choice may depend on the angle of incidence. We provide,
therefore, a mathematical prescription which converts
Eq.~(\ref{eq:atom_transampl_gra}) into an expression in the style of
the Kirchhoff integral.
This is achieved by introducing, between every pair of adjacent bars,
a new coordinate system $(s_1,s_2)$ as depicted in
Fig.~\ref{fig:atom_scattering_grating} such that the $s_2$-axis meets
the boundaries of the two bars at their respective shadow lines. The
length of the resulting straight ``slit line'' $\mathcal S$ will be
denoted by $S_0$. We now substitute the integration variable $a_2$ of
Eq.~(\ref{eq:atom_transampl_bar}) by
\begin{displaymath}
  s_2=\frac{\Delta p_{a_2}}{\Delta
  p_{s_2}}\left(a_2\pm\frac{A}2\right)\pm\frac{S_0}2
\quad\mbox{as}\quad a_2 \lessgtr 0
\end{displaymath}
where $\Delta p_{s_2}$ denotes the momentum transfer parallel to the
$s_2$-axis. Similarly, we denote the components of the incident and
outgoing momenta with respect to the $s_1$-axis by $p'_{s_1}$ and
$p_{s_1}$, respectively. (An explicit expression for these momentum
components in terms of the geometry of the grating will be given below
in Eqs.~(\ref{eq:momentum_angles}) and (\ref{eq:momtransfk}) for the
transmission grating of Fig.~\ref{fig:geometry_bar}.)  Using the
identities
\begin{subequations}\label{eq:impulsrelationen}
\begin{eqnarray}
  \label{eq:impulssumme}
  \Delta p_{a_2}A+\Delta p_{s_2} S_0&=&\Delta p_2 d\ ,\\
  \label{eq:impulsverh}
  \frac{p_{a_1}+p'_{a_1}}{\Delta p_{a_2}}=
  \frac{p_{s_1}+p'_{s_1}}{\Delta p_{s_2}}&=&
  \frac{p_1+p'_1}{\Delta p_2}\ ,
\end{eqnarray}
\end{subequations}
which hold because of energy conservation, and the abbreviation
$D=d\Delta p_2/\Delta p_{s_2}$, the transition amplitude of the single
bar Eq.~(\ref{eq:atom_transampl_bar}) becomes
\begin{align}
  \nonumber
  &t^{\text{at}{(2)}}(\vecz p;\vecz p')=
  -\frac{i}{2}\frac{p_{s_1}+p'_{s_1}}{(2\pi)^2m\hbar}
  \left\{\
  e^{ i \Delta p_2 d/2\hbar}\int_{{S_0}/2}^{{D}/2} d s_2\ 
  e^{- i\Delta p_{s_2} s_2/\hbar}
  \right.
  \\
  \label{eq:streuampl_atom_attr5}
  & + \left.
    e^{- i \Delta p_2 d/2\hbar}\int_{-{D}/2}^{-{S_0}/2} d s_2\ 
    e^{- i\Delta p_{s_2} s_2/\hbar}\
  \right\}.\hspace{0.5cm}
\end{align}
Keeping in mind that the integration variable $s_2$ was substituted
for $a_2$ for a single bar the following geometrical interpretation is
possible: in the integral from $S_0/2$ to $D/2$ in
Eq.~(\ref{eq:streuampl_atom_attr5}) the variable $s_2$ represents the
position along the upper half of the slit line $\mathcal{S}$ on one
side of the bar; similarly, in the integral running from $-D/2$ to
$-S_0/2$, $s_2$ is the position along the lower half of the next slit
line at the other side the bar.  Inserting
Eq.~(\ref{eq:streuampl_atom_attr5}) into
Eq.~(\ref{eq:atom_transampl_gra_a}) the half slit lines of the $N$
adjacent bars can be joined to yield $N-1$ slits between them.
Collecting all terms and introducing a ``slit function'',
\begin{equation}
  \label{eq:atom_slitfunction}
  a^{\text{at}}(\vecz p'; \Delta p_{s_2})=
  \int_{-{D}/2}^{{D}/2} d s_2\ 
  \exp\left(- i \Delta p_{s_2} s_2/\hbar\right)\ 
  \tau^\text{at}(\vecz p'; s_2),
\end{equation}
where the transmission function $\tau^\text{at}(\vecz p'; s_2)$
inserted here is unity inside the slit $\mathcal S$ and zero
otherwise, the transition amplitude of the grating can be written, for
$N\gg1$, as
\begin{eqnarray}
  \label{eq:atom_transampl_gra2}
  \lefteqn{t^{\text{at}{(2)}}_\text{gra}(\vecz p;\vecz p')\simeq
    -\frac{i}{2}\frac{p_{s_1}+p'_{s_1}}{(2\pi)^2m\hbar}}& &\\
  \nonumber
  &\times&
  \left\{
  \
  \frac{\sin\left(\Delta p_2 N d/2\hbar\right)}
  {\Delta p_{s_2}/2\hbar}
  -
  H_{N-1}(\Delta p_2)\ a^{\text{at}}(\vecz p'; \Delta p_{s_2})
  \right\}.
\end{eqnarray}
The first term in curly brackets is sharply peaked about the forward
direction, and in the limit $N\rightarrow\infty$, using
Eq.~(\ref{eq:impulsverh}), it simply reduces to
$2\pi\hbar\frac{p_1}{p_{s_1}}\delta(\Delta p_2)$. The second term,
which is a product of the grating function (\ref{eq:gitterfunktion})
and the slit function (\ref{eq:atom_slitfunction}), generates the
familiar diffraction pattern of a grating \cite{footnote-HN}. The
$n$-th order principle diffraction maximum appears at the momentum
transfer
\begin{displaymath}
  \Delta p_2=\frac{n2\pi\hbar}d.
\end{displaymath}
Introducing the angle of incidence $\theta'$ such that $p'_1=|\vecz
p'|\cos\theta'$ and $p'_2=|\vecz p'|\sin\theta'$
(cf.~Fig.~\ref{fig:atom_scattering_grating}), and equivalently the
diffraction angle $\theta$ such that $p_1=|\vecz p|\cos\theta$ and
$p_2=|\vecz p|\sin\theta$, the $n$-th order is located at the angle
$\theta=\theta_n$ satisfying
\begin{equation}
  \label{eq:thetan_schr}
  \sin\theta_n=\sin\theta'+\frac{n2\pi\hbar}{|\vecz p'|d}\ .
\end{equation}
Generally, the diffraction intensities are proportional to the
scattering matrix element $|\bra{\vec p}S\ket{\vec p'}|^2$ where the
components of the outgoing momentum $\vecz p$ must be evaluated at the
angle $\theta_n$.  Inserting Eq.~(\ref{eq:atom_transampl_gra2}) into
Eq.~(\ref{eq:Smatrix}) one finds, after some algebra,
\begin{equation}
  \label{eq:atom_intensity}
  I_n=I_0
  \left(\frac{p_{s_1}+p'_{s_1}}{2 p'_{s_1}}\right)^2
  \frac{| a^{\text{at}}\left(\vecz p'; \Delta p_{s_2}\right) |^2}
  {| a^{\text{at}}\left(\vecz p'; 0\right) |^2},
\end{equation}
where the intensity $I_0$ of the zeroth diffraction order serves as a
normalization constant depending on the experimental counting rate and
where, from Eq.~(\ref{eq:atom_slitfunction}), $|
a^{\text{at}}\left(\vecz p'; 0\right) |^2=S_0^2$ (in the absence of
the van der Waals interaction considered below).

\begin{figure}[htbp]
  \centering
  \epsfig{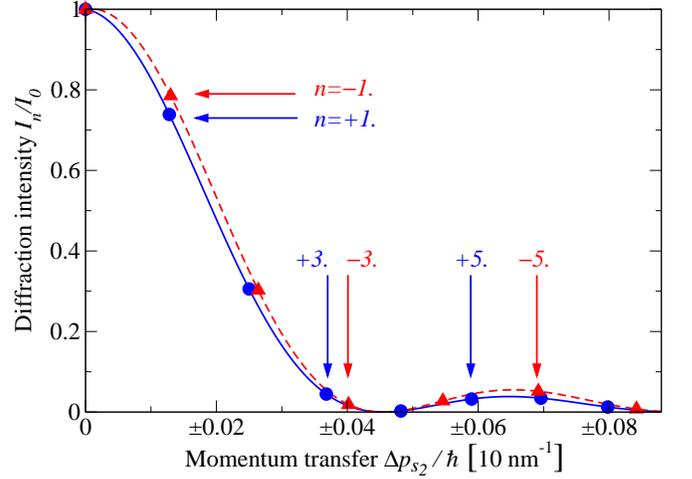}
  \caption{Asymmetric probing of the envelope function by the
    grating function at non-normal incidence. Both curves show the
    product $({p_{s_1}+p'_{s_1}})^2/({2 p'_{s_1}S_0})^2
    |a^{\text{at}}\left(\vecz p'; \Delta p_{s_2}\right)|^2$ as it
    appears in Eq.~(\ref{eq:atom_intensity}) versus the momentum
    transfer $\Delta p_{s_2}$. The solid curve refers to the positive
    values of $\Delta p_{s_2}$ on the horizontal axis. The dashed
    curve refers to the negative values of $\Delta p_{s_2}$ which have
    been mirrored onto the positive axis for comparison. The circles
    and triangles on top of the solid and the dashed curves,
    respectively, mark those values of $\Delta p_{s_2}$ where the
    grating function $H_N(\Delta p_2)$ probes the slit function,
    i.e.~where $\Delta p_2=n2\pi\hbar/d$ is satisfied for
    $n=0,\pm1,\pm2,\dots$.  Their intensities correspond to the
    measurable diffraction peaks.  For the calculation of this figure
    the grating cross section of Fig.~\ref{fig:geometry_bar} was used
    with the beam parameters $\theta'=21^\circ$ and $|\vecz
    p'|/\hbar=10$\,nm$^{-1}$.}
  \label{fig:distort}
\end{figure}
We now discuss the intensity formula Eq.~(\ref{eq:atom_intensity}) to
explain the origin of the asymmetry of the diffraction pattern.  The
slit function $a^{\text{at}}\left(\vecz p'; \Delta p_{s_2}\right)$ is
an even function of the momentum transfer component $\Delta p_{s_2}$.
The geometrical factor $({p_{s_1}+p'_{s_1}})^2/({2 p'_{s_1}S_0})^2$
(the component $p_{s_1}$ depends on $\Delta p_{s_2}$ through
conservation of energy and momentum) can be shown to introduce, for
positive incident angle $\theta'$, a slight attenuation of the slit
function at positive $\Delta p_{s_2}$ and likewise an intensification
at negative $\Delta p_{s_2}$.  The product of the slit function and
the geometrical factor serves in Eq.~(\ref{eq:atom_intensity}) as an
envelope function which is probed by the grating function at the
momentum transfer $\Delta p_2$ rather than at $\Delta p_{s_2}$.  Since
$\Delta p_{s_2}$ and $\Delta p_2$ are not proportional to each other
this probing is not symmetric for positive and negative diffraction
angles.  This is depicted in Fig.~\ref{fig:distort}, and it leads to
the characteristic asymmetry of the diffraction pattern of a deep
grating at non-normal incidence. Expanding $\Delta p_{s_2}$, using
Eq.~(\ref{eq:impulsverh}), into a power series in $(\Delta
p_2/\cos\theta')$ through second order,
\begin{align}
  \label{eq:dps2-series}
  \frac{\Delta p_{s_2}}{\cos(\alpha+\theta')}\simeq
  \frac{\Delta p_2}{\cos\theta'}+
  \frac{\tan\theta'-\tan(\alpha+\theta')}{2 |\vecz p'|}
  \left(\frac{\Delta p_2}{\cos\theta'}\right)^2,
\end{align}
the leading non-linear term is seen to vanish like $\Delta p_2/|\vecz
p'|$ relative to the linear term. Accordingly, the asymmetry is less
pronounced for smaller diffraction orders, for faster beams and, in
the case of molecules, for heavier molecules. While at $|\vecz
p'|/\hbar=10\,\text{nm}^{-1}$ (corresponding to a {\Heatom} beam at
$v\approx160\,$m/s), as seen in Fig.~\ref{fig:distort}, the quadratic
term in Eq.~(\ref{eq:dps2-series}) is responsible for a $\pm8\%$
deviation of the positive and negative fifth diffraction orders,
respectively, its contribution reduces to $\pm0.7\%$ at $v=1800\,$m/s.
This smallness explains why the asymmetry has previously been missed
(cf.~Fig.~5 in Ref.~\cite{GSTMSS_PRA61}).  Clearly, in the thin
grating limit $\alpha\rightarrow 0$ the mirror symmetry is recovered
in Eq.~(\ref{eq:dps2-series}).

In the derivation so far no comment has been made about the inclusion
of the attractive van der Waals interaction between the atom and the
grating.  It can be accounted for through the transmission function
$\tau^\text{at}(\vecz p'; s_2)$ in the slit function
(\ref{eq:atom_slitfunction}) as outlined in Appendix
\ref{sec:surface_interaction} and in
Refs.~\cite{HK_PRA61,GSTHK_PRL83}. Unlike the case of normal
incidence, at non-normal incidence the influence of the van der Waals
interaction may be different on each side of the slit, introducing, in
principle, an additional source of asymmetry to the diffraction
pattern.  Numerical comparisons using the explicit expression of
Appendix \ref{sec:surface_interaction} for the transmission function
demonstrate, however, that this effect is minor.

\subsection{The atom diffraction pattern of the deep grating}

\begin{figure}[htbp]
  \centering
  \epsfig{file=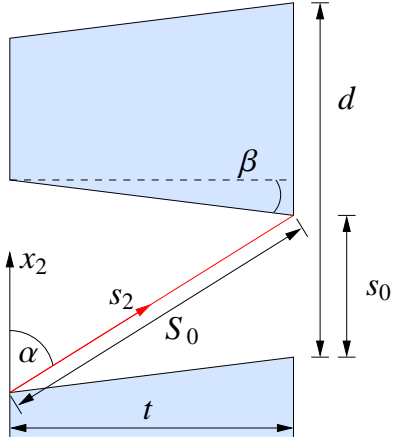,width=0.6\columnwidth}
  \caption{Geometrical cross section of a custom
    diffraction grating as it has been used in matter diffraction
    experiments \cite{GSTHK_PRL83,GSTHKS_PRL85}. The parameters and
    their typical values are: period $d=100$\,nm, slit width
    $s_0=60$\,nm, thickness $t=120$\,nm, and wedge angle
    $\beta=6^\circ$. The angle by which the coordinate systems
    $(x_1,x_2)$ and $(s_1,s_2)$ are rotated with respect to each other
    satisfies $\cot\alpha=\tan\beta+s_0/t$ and $S_0\sin\alpha=t$. At
    the above parameters $\alpha\approx58^\circ$.  The characteristic
    shape of the bars is reminiscent of the lithographic production
    process \cite{SSCS_JVSTB13}.}
  \label{fig:geometry_bar}
\end{figure}

\begin{figure}[htbp]
  \centering
  \epsfig{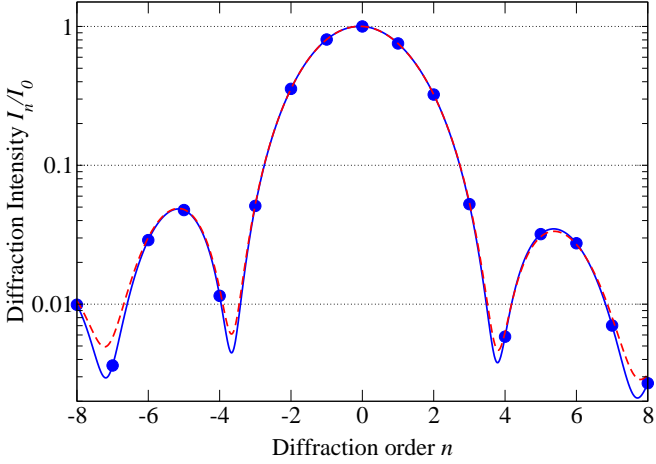}
  \caption{Diffraction intensities of a
    Helium atom beam at $v=500$\,m/s from a $d=100$\,nm transmission
    grating as displayed in Fig.~\ref{fig:geometry_bar} at
    $\theta'=21^\circ$ angle of incidence.  The solid curve was
    calculated using Eq.~(\ref{eq:atom_intensity}), the dashed curve
    shows the two-term cumulant approximation
    (\ref{eq:atom_intensity_cumul}). To guide the eye these functions
    are shown continuously. The circles on top of the solid curve at
    integer $n$ mark the experimentally accessible diffraction orders
    $I_n/I_0$.}
  \label{fig:In_Atom_Cumulants}
\end{figure}

The quantitative evaluation of experimental diffraction data requires
to determine a set of parameters describing the geometry of the
particular grating (cf.~Fig.~\ref{fig:geometry_bar}) as well as the
van der Waals interaction coefficient $C_3$ (cf.~appendix
\ref{sec:surface_interaction}).
Previous work has shown that an immediate numerical fit of
Eq.~(\ref{eq:atom_intensity}) to experimental data does not reliably
determine these parameters. Analogous to the procedure developed in
Ref.~\cite{GSTHK_PRL83} for diffraction at normal incidence we
therefore introduce a two-term cumulant approximation of the slit
function. To this end we rewrite Eq.~(\ref{eq:atom_slitfunction}) as
\begin{align}
  \nonumber
  &a^{\text{at}}({\vecz p}'; \Delta p_{s_2})=
  \frac{\hbar}{ i \Delta p_{s_2}} 
  \\
  \label{eq:atom_slitfunction_charfct}
  &\times\ 
  \left\{
  e^{ i \Delta p_{s_2} S_0/2\hbar}\
  \Phi^{-}\left(\frac{\Delta p_{s_2}}\hbar\right)
  -e^{- i \Delta p_{s_2} S_0/2\hbar}\
  \Phi^{+}\left(\frac{\Delta p_{s_2}}\hbar\right)
  \right\}
\end{align}
where the functions $\Phi^+(\kappa)$ and $\Phi^-(\kappa)$ are defined by
\begin{equation}
  \label{eq:charFkt}
  \Phi^\pm(\kappa)=
  \frac{\mp 1}{\tau^\text{at}\left({\vecz p}'; 0\right)}
  \int_0^{{S_0}/2} d\xi\ e^{\pm i\kappa\xi}\
  {\tau^\text{at}}'\left({\vecz p}'; \pm\left(\frac{S_0}2-\xi\right)\right).
\end{equation}
Here, ${\tau^\text{at}}'(\vecz p'; s_2)$ denotes the derivative of the
transmission function with respect to its position argument. As
$\Phi^\pm(0)=1$ the logarithm of $\Phi^\pm(\kappa)$ can be expanded
into a power series of the form
\begin{equation}
  \label{eq:logcharFkt}
  \ln \Phi^\pm(\kappa)=
  \sum_{j=1}^\infty \frac{(\pm i\kappa)^j}{j!} R^\pm_j.
\end{equation}
The complex numbers $R^\pm_1$ and $R^\pm_2$, which are known as the
first two cumulants, are uniquely determined by
Eqs.~(\ref{eq:charFkt}) and (\ref{eq:logcharFkt}). One finds
\begin{eqnarray}
  \label{eq:R1pm}
  R^\pm_1
  &=&\frac{S_0}2-\int_0^{{S_0}/2} d\xi\ 
  \tau^\text{at}\left({\vecz p}';
    \pm\left(\frac{S_0}2-\xi\right)\right)
\end{eqnarray}
and 
\begin{align}
  \label{eq:R2pm}
  R^\pm_2=
  \left(\frac{S_0}2\right)^2-\left(R^\pm_1\right)^2
  -2 \int_0^{S_0/2}d\xi\ \xi\
  \tau^\text{at}\left({\vecz p}'; \pm\left(\frac{S_0}2-\xi\right)\right) .
\end{align}
Using the explicit form of the transmission function derived in the
appendix the length scale of the cumulants can be shown to be set by
the parameter $\sqrt{C_3/(\hbar v)}$. For helium and a SiN$_x$ grating
$C_3\approx 0.1\,$meV nm$^3$ \cite{GSTHK_PRL83}.  Therefore, for the
purposes of this work it is sufficient to truncate the expansion
Eq.~(\ref{eq:logcharFkt}) after the second order. Inserting the first
two terms into the slit function
Eq.~(\ref{eq:atom_slitfunction_charfct}) and introducing the four
quantities
\begin{align}
  \nonumber
  &S_\text{eff}=S_0-\text{\,Re\,}(R^+_1+R^-_1), 
  &\Delta=\text{\,Im\,}(R^+_1+R^-_1),\\
  \nonumber
  &\Gamma=\text{\,Im\,}(R^+_1-R^-_1),
  &\Sigma=\sqrt{\frac12\text{\,Re\,}(R^+_2+R^-_2)},
\end{align}
the $n$-th order diffraction intensity relative to the zeroth order is
given, within this approximation, by
\begin{align}
  \nonumber
  \frac{I_n}{I_0}=&\left[\frac{p_{s_1}+p'_{s_1}}
    {2 p'_{s_1} \sqrt{S_\text{eff}^2+\Delta^2}}\right]^2
  \exp\left[-(\Delta p_{s_2}\Sigma)^2/\hbar^2\right]\ 
  \exp[-\Gamma \Delta p_{s_2}/\hbar]\\
  \label{eq:atom_intensity_cumul}
  &\ \times\ 
  \frac{\sin^2[\Delta p_{s_2}S_\text{eff}/2\hbar]+
    \sinh^2[\Delta p_{s_2} \Delta /2\hbar]}
  {\left[\Delta p_{s_2}/2\hbar\right]^2}.
\end{align}
Here, the momentum components are to be taken explicitly at the angles
\begin{subequations}
  \label{eq:momentum_angles}
  \begin{align}
    &p_{s_1}=|\vecz p'|\cos(\alpha+\theta_n)\\
    &p'_{s_1}=|\vecz p'|\cos(\alpha+\theta')
  \end{align}
\end{subequations}
and the momentum transfer is given by 
\begin{align}
  \label{eq:momtransfk}
  \Delta p_{s_2}=|\vecz p'|\left[\sin(\alpha+\theta_n) 
    -\sin(\alpha+\theta')\right]
\end{align}
where $\alpha$ denotes the angle shown in Fig.~\ref{fig:geometry_bar}
by which the coordinate system $(s_1,s_2)$ is rotated with respect to
$(x_1,x_2)$. The diffraction intensity formula
(\ref{eq:atom_intensity_cumul}) is, though more complicated,
reminiscent of the result for normal incidence derived in
Ref.~\cite{GSTHK_PRL83}: The long fraction involving the sine of
$S_\text{eff}$ resembles the Kirchhoff slit function for a slit of
``effective width'' $S_\text{eff}$ whose intensity zeros, though, are
removed by the hyperbolic sine function involving $\Delta$. The
Gaussian exponential reflects the suppression of higher diffraction
orders. The asymmetry $I_n\neq I_{-n}$ of the diffraction pattern is
now embodied by the asymmetry in $\Delta p_{s_2}$ as a function of $n$
in Eq.~(\ref{eq:momtransfk}).  The second exponential in
Eq.~(\ref{eq:atom_intensity_cumul}) accounts for the minor additional
asymmetry in the diffraction pattern due to the different influence of
the van der Waals interaction on both sides of the bar. A comparison
of the diffraction intensities calculated from
Eq.~(\ref{eq:atom_intensity}) and the approximation
(\ref{eq:atom_intensity_cumul}) is displayed in
Fig.~\ref{fig:In_Atom_Cumulants}.

By a numerical fit of Eq.~(\ref{eq:atom_intensity_cumul}) to
experimental diffraction data the effective slit width $S_\text{eff}$,
amongst the other parameters, can be determined accurately and allows
for further comparison between theory and experiment along the lines
of Ref.~\cite{GSTHK_PRL83}. For
completeness, we note that
\begin{equation}
  \label{eq:Seff-1}
  S_\text{eff}=S_0-\text{\,Re\,}\int_{-{S_0}/2}^{{S_0}/2} d s_2
  \left[1-\tau^\text{at}({\vecz p}'; s_2) \right].
\end{equation}
This means that the geometrical slit width $S_0$ appears to be reduced
by the average deviation from unity of the transmission function
$\tau^\text{at}({\vecz p}'; s_2)$.

\section{Trimer diffraction theory\label{sec:trimerscatteringtheory}}

A trimer, in the scope of this article, is a three-atomic molecule
which is weakly bound by pair interactions
\cite{footnote-othertrimers}.  Again, the atoms themselves are treated
as point particles.  An additional three-body interaction between the
atoms is assumed to be negligible.  Central to later applications will
be the helium trimer \Hetrimer\ in which case at least these
assumptions are expected to be valid \cite{RA_JCP102}.

\subsection{Three-body bound states}

The masses of the three atoms at positions $\vec r_i$, for $i=1,2,3$,
will be denoted by $m_i$ and are assumed to be of the same order of
magnitude. The interaction between atom $j$ and $k$ is modeled by a
potential $v_{jk}(|\vec r^{(jk)}|)$ where $\vec r^{(jk)}=\vec
r_j-\vec r_k$ is the relative coordinate.
\begin{figure}[htbp]
  \centering
  \epsfig{file=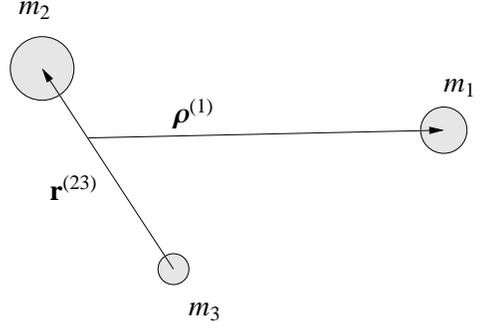,width=0.8\columnwidth}
  \caption{One of three possible sets of Jacobi
    coordinates. The vector $\vec\rho^{(1)}$ points from the center of
    mass of the subsystem $(23)$ to atom 1. The vector $\vec r^{(23)}$
    is the relative coordinate of the subsystem. The coordinate $\vec
    R$ (not shown) corresponds to the center of mass position and is,
    therefore, identical for all three sets.}
  \label{fig:jacobi-coordinates}
\end{figure}
We introduce the Jacobi coordinates $\vec R,\vec \rho^{(i)}, \vec
r^{(jk)}$ sketched in Fig.~\ref{fig:jacobi-coordinates}, which can be
expressed in block matrix form as
\begin{equation}
  \label{eq:jacobi-coordinates-pos}
  \left( \begin{array}{l}
      \vec R\\ \vec \rho^{(i)}\\ \vec r^{(jk)}
    \end{array} \right) =
  \left( \begin{array}{rrr}
      \frac{m_i}M \mathbf{1} & \frac{m_j}M \mathbf{1} &
      \frac{m_k}M \mathbf{1}\\
      \mathbf{1} & -\frac{m_j}{m_j+m_k} \mathbf{1} &
      -\frac{m_k}{m_j+m_k}\mathbf{1} \\
      \mathbf{0} & \mathbf{1} & -\mathbf{1}
    \end{array} \right)
  \left( \begin{array}{l}
      \vec r_i\\ \vec r_j\\ \vec r_k
    \end{array} \right) ,
\end{equation}
where $\mathbf{1}$ and $\mathbf{0}$ denote the $3\times3$ unit and
zero matrix, respectively, and $M=m_1+m_2+m_3$ is the total mass. It
is sufficient to restrict the combinations of indices to the ascending
permutations
\begin{equation}
  \label{eq:ijk_trimerindices}
  (ijk)=(123),(231),(312).
\end{equation}
The transformation between different sets of Jacobi coordinates can be
derived from Eq.~(\ref{eq:jacobi-coordinates-pos}). It takes the form
\begin{equation}
  \label{eq:jacobi-coordinates-pos-transf}
  \left( \begin{array}{l}
      \vec R\\ \vec \rho^{(j)}\\ \vec r^{(ki)}
    \end{array} \right) =
  \mathcal{J}^{(ji)}
  \left( \begin{array}{l}
      \vec R\\ \vec\rho^{(i)}\\ \vec r^{(jk)}
    \end{array} \right),
\end{equation}
where the block matrix $\mathcal{J}^{(ji)}$ is given by
\begin{equation}
  \label{eq:jacobi-relkoordinaten-ort-transf-Dji}
  \mathcal{J}^{(ji)}=
  \left( \begin{array}{rrr}
      \mathbf{1} & \mathbf{0} & \mathbf{0}\\
      \mathbf{0} & -\frac{m_i}{m_i+m_k} \mathbf{1} &
      \frac{m_k M}{(m_j+m_k)(m_k+m_i)} \mathbf{1}\\
      \mathbf{0} & 
      -\mathbf{1} & -\frac{m_j}{m_j+m_k} \mathbf{1}
    \end{array} \right).
\end{equation}
The three matrices $\mathcal{J}^{(ji)}$ satisfy the relations $\det
\mathcal{J}^{(ji)} = 1$ and
$\mathcal{J}^{(ji)}\mathcal{J}^{(ik)}\mathcal{J}^{(kj)}= \mathbf{1}$.
Expressed in Jacobi coordinates the Hamilton operator for a free
trimer is given by $H_0+V$ where
\begin{align}
  \nonumber
  H_0=&\ \frac{1}{2M}\vec{\hat P}^2
  +\frac{M}{2 m_i(m_j+m_k)}{\vec{\hat q}^{(i)\,2}}
  +\frac{m_j+m_k}{2m_j m_k}{\vec{\hat p}^{(jk)\,2}}\\
  \nonumber
  V=\ &
    v_{ij}\left(\left|
        \hat{\vec\rho}^{(i)}-
        \frac{m_k}{m_j+m_k}\vec{\hat r}^{(jk)}\right|\right)
    +v_{jk}\left(\left|\vec{\hat r}^{(jk)}\right|\right)\\
  \label{eq:hamilton_trimer}
  &+\ 
  v_{ki}\left(\left|
      \hat{\vec\rho}^{(i)}+
      \frac{m_j}{m_j+m_k}\vec{\hat r}^{(jk)}\right|\right).
\end{align}
Here, $\vec P,\vec q^{(i)}$, and $\vec p^{(jk)}$ are the conjugate
momenta associated with $\vec R,\vec \rho^{(i)}$, and $\vec r^{(jk)}$,
respectively. Denoting the eigenstates of the center of mass momentum
$\hat{\vec P}$ by $\ket{\vec P}$ the full trimer states can be written
in product form $\ket{\vec P,\phi_{\gamma}}\equiv\ket{\vec
  P}\ket{\phi_{\gamma}}$ satisfying
\begin{equation}
  \label{eq:hamilton_trimer_seq}
  [H_0+V]\ket{\vec P,\phi_{\gamma}}=
  E\ket{\vec P,\phi_{\gamma}}
\end{equation}
with energy eigenvalues
\begin{equation}
  \label{eq:energie_trimer}
  E=\frac{|\vec P|^2}{2M}+E_{\gamma}
\end{equation}
where $E_{\gamma}$ is the negative binding energy of the trimer bound
state $\ket\phi_{\gamma}$.

The representation of a trimer state by its wave function depends on
the particular set of Jacobi coordinates.  Denoting the common
eigenstates of the relative momentum operators $\vec{\hat q}^{(i)}$
and $\vec{\hat p}^{(jk)}$ by $\ket{\vec q,\vec p}_{i,jk}$, where $\vec
q$ and $\vec p$ are the corresponding eigenvalues, we introduce
momentum space wave functions by
\begin{equation}
  \label{eq:wavefunctions_ijk}
  \phi_{\gamma}^{(i,jk)}(\vec q,\vec p)=
  \ _{i,jk}\braket{\vec q,\vec p}{\phi_{\gamma}}. 
\end{equation}
Because of the transformation
Eq.~(\ref{eq:jacobi-coordinates-pos-transf}) wave functions with
respect to different sets of Jacobi coordinates can be chosen to
satisfy the transformation relation
\begin{equation}
  \label{eq:wavefunctions_ijk_jki}
  \phi_{\gamma}^{(i,jk)}(\vec q^{(i)},\vec p^{(jk)})=
  \phi_{\gamma}^{(j,ki)}(\vec q^{(j)},\vec p^{(ki)}).
\end{equation}
The corresponding configuration space wave functions
$\phi_{\gamma}^{(i,jk)}(\vec \rho,\vec r)$ are defined analogously.
In order not to overload the notation we will omit in the following,
where possible, the indices of the relative coordinates: if not
denoted otherwise we implicitly use $(ijk)=(123)$ whence
$\vec\rho\equiv\vec\rho^{(1)}$, $\vec r\equiv\vec r^{(23)}$ and $\vec
q\equiv\vec q^{(1)}$, $\vec p\equiv\vec p^{(23)}$.

\begin{figure}[htbp]
  \centering
  \epsfig{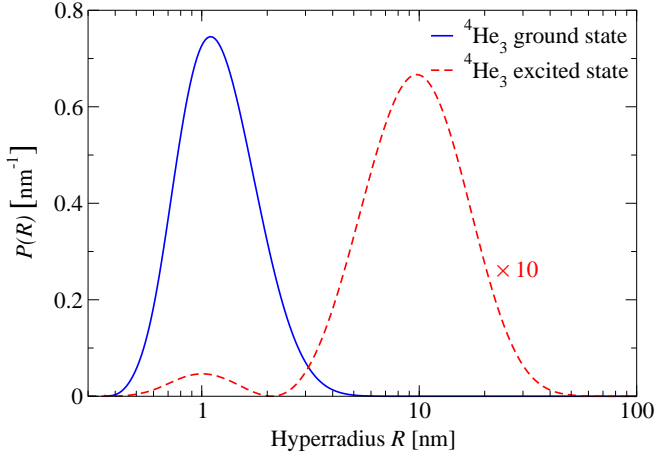}
  \caption{Hyperradial probability densities $P(R)$
    according to Eq.~(\ref{eq:hyperPR}) of the two theoretically
    predicted bound states of the helium trimer {\Hetrimer}. The
    states were calculated numerically using the momentum space
    Faddeev equations \cite{F_JETP12,sitenko} in the unitary pole
    approximation \cite{L_PR135} based on the TTY potential
    \cite{TTY_PRL74} for the helium-helium interaction. Clearly, the
    excited state, with its expectation value of the hyperradius of
    $\sqrt{\mu_0/m}\expect{R}_\text{e}=10.1\,$nm, is spatially more
    extended by almost one order of magnitude than the ground state
    with $\sqrt{\mu_0/m}\expect{R}_\text{g}=1.1\,$nm. The scaling of
    the horizontal axis is logarithmic.}
  \label{fig:heliumtrimer_PR}
\end{figure}
Since the discovery of the Efimov effect \cite{E_PL33B} in 1970 the
helium trimer {\Hetrimer} has received much attention \cite{LDD_PRL38,
  HL_JCP68,NL_PRA28,FJ_PRL71,NFJ_JPB31,MSK_CPL275,KMS_JPB31,RY_CPL328,MSSK_JPB31,ELG_PRA54,B_JCP110,BK_PRA64,L_JCP106,RA_JCP102,BH_PRA67-1,BH_PRA67-2,YMTF_PRA68,HK_PRL84}.
This trimer is predicted to possess, apart from its quite tightly
bound ground state ($E_\text{g}=-126$\,mK), a weakly bound
Efimov-type excited state ($E_\text{e}=-2.3$\,mK) \cite{BK_PRA64}.
As both states have zero total angular momentum the corresponding wave
functions only depend on three coordinates which may be taken as
$\rho=|\vec\rho|, r=|\vec r|$, and the angle between $\vec\rho$ and
$\vec r$. A common way to visualize trimer wave functions is to draw
the hyperspherical probability density $P(R)$: The hyperradius $R$,
which is independent of the choice of the set of Jacobi coordinates,
is defined as $\mu_0 R(\rho,r)^2=\frac23 m \rho^2+\frac12 m r^2$ and
the corresponding probability density can be calculated according to
\begin{equation}
  \label{eq:hyperPR}
  P(R)=\int d^3\rho d^3 r
  |\phi_{\gamma}(\vec\rho,\vec r)|^2 \delta(R(\rho,r)-R).
\end{equation}
The purpose of the ``mass'' parameter $\mu_0$ is to ensure that the
unit of $R$ is length. The numerical value of $\mu_0$ is, in
principle, arbitrary as it simply scales $R$.
Fig.~\ref{fig:heliumtrimer_PR} displays the hyperradial probability
densities for the two helium trimer states, using $\mu_0/m=\frac12$,
which were calculated numerically from the momentum space Faddeev
equations \cite{F_JETP12,sitenko} in the unitary pole approximation
\cite{L_PR135} based on the TTY potential \cite{TTY_PRL74}.

\subsection{Scattering theory approach to trimer diffraction}

We now proceed to the diffraction of a trimer from an external
potential
\begin{equation}
  \label{eq:potential_trimer}
  W(\vec r_1,\vec r_2,\vec r_3)=W_1(\vec r_1)+W_2(\vec r_2)+
  W_3(\vec r_3)
\end{equation}
where the $W_i(\vec r_i)$ are the interactions of the individual atoms
with the diffraction object.  The full Hamilton operator is given by
\begin{equation}
  \label{eq:hamilton}
  H=H_0+V+W.
\end{equation}
By virtue of this structure, which is formally identical to that of
dimer diffraction \cite{HK_PRA57}, we may carry over the fundamental
algebraic relations from previous work.  Introducing, as in
Ref.~\cite{HK_PRA57}, the resolvents
\begin{subequations}
  \label{eq:resolvents_trimer}
  \begin{eqnarray}
    \label{eq:G0z_trimer}
    G_0(z)&=&\left[z-H_0\right]^{-1}\\
    \label{eq:GVz_trimer}
    G_V(z)&=&\left[z-H_0-V\right]^{-1}\\
    \label{eq:GWz_trimer}
    G_W(z)&=&\left[z-H_0-W\right]^{-1}
\end{eqnarray}
\end{subequations}
and the two-body $T$ matrices in three-body space,
\begin{eqnarray}
  \label{eq:TVop_trimer}
  T_V(z)&=&V+V G_V(z) V\\
  \label{eq:TWop_trimer}
  T_W(z)&=&W+W G_W(z) W
\end{eqnarray}
for the potentials $V$ and $W$, an AGS type \cite{AGS} transition
operator $U_{VV}$ can be derived which satisfies the equation
\begin{equation}
  \label{eq:UVV}
  U_{VV}=T_W+T_W G_0 T_V G_0 U_{VV}.
\end{equation}
In particular, the transition amplitude is given by the matrix element
of this transition operator \cite{HK_PRA57},
\begin{equation}
  \label{eq:trimer_transampl}
  t(\vec P,\phi_{\gamma};\vec P',\phi_{\gamma'})=
  \bra{\vec P,\phi_{\gamma}}U_{VV}(E'+i0)\ket{\vec P',\phi_{\gamma'}},
\end{equation}
and it determines the $S$ matrix associated with $H$ as
\begin{align}
  \nonumber
  \bra{\vec P,\phi_{\gamma}}S_{VV}\ket{\vec P',\phi_{\gamma'}}=\ &
  \delta^{(3)}(\vec P-\vec P')\delta_{\gamma\gamma'}\\
  \label{eq:trimer_scatteringmatrix}
  & - 2\pi i\delta(E-E')t(\vec P,\phi_{\gamma};\vec P',\phi_{\gamma'}).
\end{align}
We shall in the following impose the condition
\begin{equation}
  \label{eq:EkinggEgp}
  |E_{\gamma}|,|\bra{\phi_{\gamma}}V\ket{\phi_{\gamma}}| \ll 
  \frac{|\vec P|^2}{2M}, \frac{|\vec P'|^2}{2M}
\end{equation}
which ensures that the internal energies of the trimer (both binding
energy and potential energy) are much smaller than the external energy
associated with the center of mass motion. For a helium trimer beam at
an incident beam velocity of $v=500$\,m/s, for example, the center of
mass kinetic energy ${|\vec P'|^2}/({2M})\approx 16\,$meV
(corresponding to $180$\,K) exceeds the trimer ground state energy by
more than three orders of magnitude.  Using the Schr{\"o}dinger
equation for bound trimer wave functions $\phi_{\gamma}(\vec q,\vec p)$ the
condition (\ref{eq:EkinggEgp}) can be shown to entail the relations
\begin{equation}
  \label{eq:trimer_pqllP}
  |\vec q|\ll|\vec P|,|\vec P'|
  \quad\mbox{and}\quad
  |\vec p|\ll|\vec P|,|\vec P'|.
\end{equation}
These state that the wave functions of the trimer are concentrated in
momentum space at relative momenta far smaller than the center of
mass momentum.

Under the conditions (\ref{eq:EkinggEgp}) and (\ref{eq:trimer_pqllP}) an
approximation of the equation for $U_{VV}$ (\ref{eq:UVV}) to lowest
order is possible and sufficient \cite[chap.~3.4]{gloeckle} whence the
transition amplitude becomes
\begin{equation}
  \label{eq:trimer_transampl1}
  t(\vec P,\phi_{\gamma};\vec P',\phi_{\gamma'})\simeq
  \bra{\vec P,\phi_{\gamma}}T_W(E'+i0)\ket{\vec P',\phi_{\gamma'}}.
\end{equation}
We note that within this approximation the trimer binding potential
$V$ is only implicitly contained through the bound states
$\ket\phi_{\gamma}$ and $\ket\phi_{\gamma'}$.  The evaluation of the
right hand side of Eq.~(\ref{eq:trimer_transampl1}) is nontrivial.  A
series of approximations, all accurate within the condition
(\ref{eq:EkinggEgp}), may however be applied to simplify the
transition amplitude. As the first step, the matrix element of
$T_W(E'+i0)$ can be shown to vary slowly under a variation of $E'$ on
the scale of the binding energies $E_{\gamma'}$. This allows to
replace the energy argument of $T_W$ in
Eq.~(\ref{eq:trimer_transampl1}) by the sum $E'_1+E'_2+E'_3$ where
$E'_i=|\vec p'_i|^2/2m_i$ are the energies of the free atoms.
Introducing two complete sets of states
Eq.~(\ref{eq:trimer_transampl1}) becomes
\begin{eqnarray}
  \nonumber
  & &\hspace{-0.6cm}
    t(\vec P,\phi_{\gamma};\vec P',\phi_{\gamma'})\simeq
    \int d^3 q\, d^3p\, d^3 q' d^3 p'\,
    \phi_{\gamma}^*(\vec q,\vec p)\phi_{\gamma'}(\vec q',\vec p')\ 
  \\
  \label{eq:trimer_transampl2}
  &\times&
    \bra{\vec p_1,\vec p_2,\vec p_3}T_W(E'_1+E'_2+E'_3+i0)
    \ket{\vec p'_1,\vec p'_2,\vec p'_3}.
\end{eqnarray}
Using Eqs.~(\ref{eq:resolvents_trimer}) and
(\ref{eq:TWop_trimer}) the algebraic relation $T_W(z)G_0(z)=W G_W(z)$
can be shown to hold. Inserting this relation the matrix element of
$T_W(E'_1+E'_2+E'_3+i0)$ is replaced by
\begin{equation}
  \label{eq:trimer_Welement}
  \bra{\vec p_1,\vec p_2,\vec p_3}W
  \ket{\vec p'_1,\vec p'_2,\vec p'_3,+}
\end{equation}
where $\ket{\vec p'_1,\vec p'_2,\vec p'_3,+}\equiv \ket{\vec
  p'_1,+}\ket{\vec p'_2,+}\ket{\vec p'_3,+}$ is the scattering state
of three independent atoms associated with the potential $W(\vec
r_1,\vec r_2,\vec r_3)$.  Splitting $W$ according to
Eq.~(\ref{eq:potential_trimer}) into the individual potentials
$W_i(\vec r_i)$ and using the Lippmann-Schwinger
Eq.~(\ref{eq:atomLSG}) the matrix element (\ref{eq:trimer_Welement})
can be expressed by the known atom transition amplitudes
(\ref{eq:atomstreuamplitude}):
\begin{align}
  \nonumber
  &\bra{\vec p_1,\vec p_2, \vec p_3}
  W\ket{\vec p'_1,\vec p'_2, \vec p'_3,+}=
  t^\text{at}_1(\vec p_1;\vec p'_1)\ \delta^{(3)}(\vec p_2-\vec p'_2)\\
  \nonumber
  &\times\ 
  \delta^{(3)}(\vec p_3-\vec p'_3) + t^\text{at}_1(\vec p_1;\vec p'_1)\
  \delta^{(3)}(\vec p_2-\vec p'_2)\ t^\text{at}_3(\vec p_3;\vec p'_3)
  \\
  \nonumber
  &\times\
  \left[\frac{1}{E'_3-E_3+i0}+\frac{1}{E'_1-E_1+i0}\right]
  +t^\text{at}_1(\vec p_1;\vec p'_1)\ t^\text{at}_2(\vec p_2;\vec p'_2)\\
  \label{eq:TW_4}
  &\times\ 
  t^\text{at}_3(\vec p_3;\vec p'_3)
  \frac{1}{E'_2-E_2+i0}\ \frac{1}{E'_3-E_3+i0}
  + \text{cycl.~perm.}
\end{align}
Here, ``cycl.~perm.'' indicates that all explicitly shown terms on the
right hand side of Eq.~(\ref{eq:TW_4}) should be repeated with their
indices permuted once and twice, in ascending order. Applying again the
condition of the weak binding energy (\ref{eq:EkinggEgp}) the complex
energy denominators can be approximated. Firstly, using the principal
value formula $(x+i0)^{-1}=- i\pi\delta(x)+\mathcal{P}x^{-1}$ it is
possible to approximate
\begin{eqnarray}
  \nonumber
  \lefteqn{\delta^{(3)}(\vec p_2-\vec p'_2)
    \left[\frac{1}{E'_3-E_3+i0}+
      \frac{1}{E'_1-E_1+i0}\right]}\hspace{1cm}& &\\ 
  \label{eq:distribapprox1}
  &\simeq&\delta^{(3)}(\vec p_2-\vec p'_2)
  \left[-2\pi i\delta\left(E_1-E'_1\right)\right]
\end{eqnarray}
where a small correction term $\mathcal{O}(E_{\gamma}/E)$ was neglected.
Secondly, for three variables $x,y,z$ with $x+y+z=0$ the distribution
identity
\begin{eqnarray}
  \nonumber
  \lefteqn{\frac1{x+i0}\frac1{y+i0}+\frac1{y+i0}\frac1{z+i0}+
    \frac1{z+i0}\frac1{x+i0}}\hspace{1mm} &&\\
  \label{eq:distribident2}
  &=&-\frac{(2\pi)^2}3\left[
    \delta(x)\delta(y)+\delta(y)\delta(z)+\delta(z)\delta(x)
  \right]
\end{eqnarray}
can be shown to hold. If $x+y+z\neq0$, such as in Eq.~(\ref{eq:TW_4})
for $x=E'_1-E_1$, $y=E'_2-E_2$, and $z=E'_3-E_3$,
Eq.~(\ref{eq:distribident2}) is still applicable within the same range
of validity as Eq.~(\ref{eq:distribapprox1}).  Combining the steps one
arrives at the following approximate expression for the matrix element
of $T_W(E'_1+E'_2+E'_3+i0)$:
\begin{align}
  \nonumber
  & \bra{\vec p_1,\vec p_2, \vec p_3}
  T_W(E'_1+E'_2+E'_3+i0)
  \ket{\vec p'_1,\vec p'_2, \vec p'_3}\simeq
  t^\text{at}_1(\vec p_1;\vec p'_1)\\
  \nonumber
  &\times\ 
  \delta^{(3)}(\vec p_2-\vec p'_2)\ 
  \delta^{(3)}(\vec p_3-\vec p'_3)
  - 2\pi i\ \delta\left(E_1-E'_1\right)
  t^\text{at}_1(\vec p_1;\vec p'_1)
  \\
  \nonumber
  &\times\ 
  \delta^{(3)}(\vec p_2-\vec p'_2)\ 
  t^\text{at}_3(\vec p_3;\vec p'_3)
  -
  \frac{(2\pi)^2}3\delta\left(E_2-E'_2\right)
  \delta\left(E_3-E'_3\right)\\
  \label{eq:TW_5}
  &\times\ 
  t^\text{at}_1(\vec p_1;\vec p'_1)\
  t^\text{at}_2(\vec p_2;\vec p'_2)\
  t^\text{at}_3(\vec p_3;\vec p'_3)
  +\text{cycl.~perm.}
\end{align}
Upon insertion of Eq.~(\ref{eq:TW_5}) into the trimer transition
amplitude (\ref{eq:trimer_transampl2}) several momentum integrals can
be carried out by virtue of the momentum delta-functions. Moreover,
the energy delta-functions allow the integration of a further momentum
component each.  As an example, we consider the delta-function
$\delta(E_1-E'_1)$ in the second term of Eq.~(\ref{eq:TW_5}).
Switching back to Jacobi coordinates, after integrating over
$\delta^{(3)}(\vec p_2-\vec p'_2)$ it becomes
\begin{eqnarray}
  \nonumber
  \lefteqn{\delta(E_1-E'_1)=\delta\left(
      \left[\frac{m_1}M\vec P+\vec q\right]^2\right.}& &\\
  \label{eq:energie_deltafunktion1_1}
  & &\left.
    -\left[\vec P'-\frac{m_2+m_3}M\vec P+\vec q-
      \frac{m_2+m_3}{m_2}\Delta\vec p\right]^2\right).
\end{eqnarray}
To proceed we decompose the momentum vectors into their components
parallel and perpendicular to the incident center of mass momentum
$\vec P'$: the parallel component of $\vec q$, for example, is denoted
by $q_\parallel$ and the two-dimensional perpendicular vector is
denoted by $\vec q_\perp$. Then, by condition (\ref{eq:trimer_pqllP})
and using $\vec P'_\perp=0$ by definition, the delta-function
(\ref{eq:energie_deltafunktion1_1}) can be approximated by
\begin{displaymath}
  \nonumber
  \delta(E_1-E'_1)\simeq
  \frac{
    \delta\left(\Delta P_\parallel+\frac{m_2+m_3}{m_2}\Delta p_\parallel
      +\xi\ (\frac{m_1}M P_\parallel+q_\parallel)
    \right)
  }
  {2\frac{m_2+m_3}{m_2}\left(\frac{m_1}M P_\parallel+q_\parallel\right)}
\end{displaymath}
where the term involving the factor $\xi\approx
6(\lambda_\text{dB}/d)^2$ is a correction which is small by the
diffraction condition (\ref{eq:diffractioncondition}). Integrating,
according to Eq.~(\ref{eq:trimer_transampl2}), the second term of
Eq.~(\ref{eq:TW_5}) over $ d p'_\parallel$, the correction involving
$\xi$ is pushed into the functional arguments of the atom transition
amplitudes as well as the trimer wave functions. In both cases,
however, these functions vary slowly on the scale of $\xi P_\parallel$
such that the correction can be shown to be negligible altogether.
Similar simplifications are readily derived for the other energy
delta-functions in Eq.~(\ref{eq:TW_5}).

Combining all steps the general expression for the trimer transition
amplitude subject to the diffraction condition and the condition of
weak binding energy is obtained. Beforehand, however, it is helpful to
introduce two abbreviations. Firstly, we express the atom transition
amplitude in terms of the momentum transfer $\Delta \vec p_i=\vec
p_i-\vec p'_i$ as
\begin{equation}
  \label{eq:atom_momtransf}
  \widetilde{t}_i(\vec p_i; \Delta \vec p_i)=t_i(\vec p_i;\vec p'_i).
\end{equation}
Secondly, we introduce a molecular ``form factor''
\begin{align}
  \label{eq:formfactor}
  F_{\gamma\gamma'}(\vec q;\vec p)=\!\!
  \int d^3 q' d^3 p'
  \phi_{\gamma}^*\left(\vec q',\vec p'\right)\
  \phi_{\gamma'}\left(\vec q+\vec q', \vec p+\vec p'\right).
\end{align}
With this notation the trimer transition amplitude assumes the form
\begin{align}
  \nonumber
  &t(\vec P,\phi_{\gamma};\vec{P}',\phi_{\gamma'})\simeq
  \ \Bigg\{ 
  F_{\gamma\gamma'}\left(
  0,-\frac{m_2+m_3}{M}\Delta \vec P_\perp;0
  \right)
  \widetilde{t}^\text{at}_1\left(\frac{m_1}M\vec P; \Delta\vec P\right)
  \\
  \nonumber
  &  -2\pi i\ \frac{M(m_2+m_3)}{m_2 P_\parallel}
  \int d^2 p_\perp \
  \widetilde{t}^\text{at}_3\left(\frac{m_3}M\vec P; 
  \Delta P_\parallel, -\frac{m_2+m_3}{m_2}\vec{p}_\perp\right)
  \\
  \nonumber
  &
  \times\ F_{\gamma\gamma'}\left(0, -\frac{m_2+m_3}{m_2}
  \left(\frac{m_2}M\Delta \vec P_\perp+\vec{p}_\perp\right);
  0,-\vec{p}_\perp\right)\\
  \nonumber
  & \times\ 
  \widetilde{t}^\text{at}_1\left(\frac{m_1}M\vec P; 
  0, \Delta \vec P_\perp+\frac{m_2+m_3}{m_2}\vec{p}_\perp\right)
  -\frac{4\pi^2}{3}\frac{M^2}{P^2_\parallel}
  \int d^2 q_\perp  d^2 p_\perp\ 
  \\
  \nonumber
  &
  \times\ 
  F_{\gamma\gamma'}\left(0,-\vec{q}_\perp;0,-\vec{p}_\perp\right)
  \widetilde{t}^\text{at}_1\left(\frac{m_1}M\vec P; 
  \Delta P_\parallel, \frac{m_1}M\Delta \vec P_\perp+
  \vec{q}_\perp\right)
  \\
  \nonumber
  &
  \times\ 
  \widetilde{t}^\text{at}_2\left(\frac{m_2}M\vec P; 
  0, \frac{m_2}M\Delta \vec P_\perp-\frac{m_2}{m_2+m_3}\vec{q}_\perp
  +\vec{p}_\perp\right)
  \\
  &\times\
  \widetilde{t}^\text{at}_3\left(\frac{m_3}M\vec P; 
    0, \frac{m_3}M\Delta \vec P_\perp-\frac{m_3}{m_2+m_3}\vec{q}_\perp
    -\vec{p}_\perp\right)
  \Bigg\}
  + \text{cycl.~perm.}
  \label{eq:TW_8}
\end{align}
where again use has been made of condition (\ref{eq:trimer_pqllP}) to
simplify the functional arguments of the atom transition amplitudes
where possible.

\section{Trimer diffraction from a deep
  grating\label{sec:trimertiltedgrating}}

\subsection{The trimer slit function of the deep grating}

In this section the general trimer transition amplitude will be
evaluated for diffraction from a deep grating.  Inserting, to this
end, the expression (\ref{eq:atom_transampl_gra2}) for the atom
transition amplitude into Eq.~(\ref{eq:TW_8}) yields a very long sum
containing a total of 42 terms which we do not spell out explicitly.
These 42 terms can be classified by the number (zero through three) of
references to the slit function in each.  In the context of a
many-body multiple scattering series expansion \cite{gloeckle} they
may respectively be interpreted as forward (9 terms), single (18
terms), double (12 terms), and triple-scattering terms (3 terms).  A
cumbersome and little elucidating calculation involving the
transformation properties of the form factor (\ref{eq:formfactor})
with respect to different sets of Jacobi coordinates reveals that the
eighteen single-scattering terms interfere almost destructively and
that their net contribution is small by a factor
$\mathcal{O}(\lambda_\text{dB}/d)^2$ compared to the forward and the
triple-scattering terms. They may thus be neglected by the diffraction
condition (\ref{eq:diffractioncondition}). Similarly, the twelve
double-scattering terms contribute only to order
$\mathcal{O}(\lambda_\text{dB}/d)$ and may also be neglected.

The forward and triple-scattering terms, respectively, can be
combined. As in the atom case in
Eq.~(\ref{eq:atomstreuamplitude3D_2D}) a delta-function $\delta(\Delta
P_3)$ can be extracted from the trimer transition amplitude leaving
\begin{equation}
  \label{eq:trimer_transampl3D_2D}
  t(\vec P,\phi_{\gamma};\vec{P}',\phi_{\gamma'})=\delta(\Delta P_3)
  t^{(2)}(\vecz P,\phi_{\gamma};\vecz{P}',\phi_{\gamma'})
\end{equation}
where bold italic letters again denote the two-dimensional projections
of vectors into the plane perpendicular to the $x_3$-axis. 
The two-dimensional trimer transition amplitude for diffraction from a
transmission grating becomes
\begin{widetext}
\begin{align}
  \nonumber
  t^{(2)}_\text{gra}&(\vecz P,\phi_{\gamma};\vecz{P}',\phi_{\gamma'})\simeq
  -\frac{i}{2}\frac{P_{s_1}+P'_{s_1}}{(2\pi)^2M\hbar}
  \ \frac13\Bigg\{ 2\pi\hbar\frac{P_1}{P_{s_1}}
  \delta(\Delta P_2)\delta_{\gamma\gamma'}
  -\frac{1}{(2\pi\hbar)^2}\frac{P^2_{s_1}}{P_\parallel^2}
  \int d q_\perp d p_\perp\ 
  F_{\gamma\gamma'}(0,-q_\perp,0;0,-p_\perp,0)\\
  \nonumber
  &\times\ 
  H_N\left(\Delta P_\parallel\sin\theta'+
    \left(\frac{m_1}{M}\Delta P_\perp+q_\perp\right)\cos\theta'\right)
  a^{\text{at}}_{1}\left(
    \frac{m_1}M \vecz P';
    \Delta P_\parallel\sin(\alpha+\theta')+
    \left(\frac{m_1}{M}\Delta P_\perp+q_\perp\right)\cos(\alpha+\theta')
  \right)
  \\
  \nonumber
  &\times\ 
  H_N\left(\left(\frac{m_2}{M}\Delta P_\perp
      -\frac{m_2}{m_2+m_3}q_\perp+p_\perp\right)\cos\theta'\right)
  a^{\text{at}}_{2}\left(
    \frac{m_2}M \vecz P';
    \left(\frac{m_2}{M}\Delta P_\perp
      -\frac{m_2}{m_2+m_3}q_\perp+p_\perp\right)\cos(\alpha+\theta')
  \right)
  \\
  \nonumber
  &\times\
  H_N\left(\left(\frac{m_3}{M}\Delta P_\perp
      -\frac{m_3}{m_2+m_3}q_\perp-p_\perp\right)\cos\theta'\right)
  a^{\text{at}}_{3}\left(
    \frac{m_3}M \vecz P';
    \left(\frac{m_3}{M}\Delta P_\perp
      -\frac{m_3}{m_2+m_3}q_\perp-p_\perp\right)\cos(\alpha+\theta')
  \right) \Bigg\}
  \\
  \label{eq:TW_10}
  &+ \text{cycl.~perm.}
\end{align}
\end{widetext}
Inserting now for each atom the slit functions (\ref{eq:atom_slitfunction})
and rewriting the form factor (\ref{eq:formfactor}) as a configuration
space integral
\begin{displaymath}
  F_{\gamma\gamma'}(\vec q;\vec p)=
  \int d^3 \rho\, d^3 r\,
  e^{- i(\vec q\cdot\vec\rho+\vec p\cdot\vec r)/\hbar}\
  \phi_{\gamma}^*(\vec \rho,\vec r)\,
  \phi_{\gamma'}\left(\vec \rho,\vec r\right)
\end{displaymath}
the integrations over $ d q_\perp$ and $ d p_\perp$ in
Eq.~(\ref{eq:TW_10}) can be carried out. The three grating functions
$H_N$ give rise to a triple sum of which only the on-diagonal terms
contribute significantly: they represent diffraction of all three
atoms from the same bar; the off-diagonal terms, which correspond to
diffraction of atoms from different bars, are negligible since the
probability for atoms to be spatially separated as far as the distance
between two adjacent bars (100\,nm) is strongly suppressed by the
bound state wave functions of the trimer.  Collecting the remaining
terms the trimer transition amplitude can be cast into the form
\begin{eqnarray}
  \label{eq:trimer_transampl3}
  \lefteqn{t^{(2)}_\text{gra}(\vecz P,\phi_{\gamma}; 
    \vecz{P}',\phi_{\gamma'})\simeq
    -\frac{i}{2}\frac{P_{s_1}+P'_{s_1}}{(2\pi)^2M\hbar}
  }& &\\
  \nonumber
  &\times&
  \Bigg\{ 2\pi\hbar\frac{P_1}{P_{s_1}}
  \delta(\Delta P_2)\delta_{\gamma\gamma'} - H_N(\Delta P_2)
  a^\text{tri}_{{\gamma\gamma'}}(\vecz P'; \Delta P_{s_2}) \Bigg\} .
\end{eqnarray}
Here, we introduced a trimer slit function by 
\begin{align}
  \label{eq:trimer_slitfunction}
  a^\text{tri}_{{\gamma\gamma'}}(\vecz P'; \Delta P_{s_2})=
  \int_{-{D}/2}^{{D}/2} d S_2
  \exp\left(- i\Delta P_{s_2} S_2/\hbar\right)
  \tau^\text{tri}_{{\gamma\gamma'}}(\vecz P'; S_2)
\end{align} 
where $S_2$ can be interpreted geometrically as the center of mass
position of the trimer along the slit line
(cf.~Fig.~\ref{fig:atom_scattering_grating}) and where, analog to the
atom case, $D=d\Delta P_2/\Delta P_{s_2}$.  Both
Eqs.~(\ref{eq:trimer_transampl3}) and (\ref{eq:trimer_slitfunction})
exhibit the same structure as their atom counterparts. Only the new
trimer transmission function, which appears in the trimer slit
function (\ref{eq:trimer_slitfunction}), and which turns out as
\begin{align}
  \nonumber
  &\tau^\text{tri}_{{\gamma\gamma'}}(\vecz P'; S_2)= \int d^3\rho\, d^3 r\ 
  \phi_{\gamma}^*(\vec \rho,\vec r)\
  \phi_{\gamma'}\left(\vec \rho,\vec r\right)\\
  \nonumber
  &\times
  \tau^\text{at}_{1}\left(\frac{m_1}M\vecz P';
    \frac{r_{1\perp}}{\cos(\alpha+\theta')}\right)
  \tau^\text{at}_{2}\left(\frac{m_2}M\vecz P';
    \frac{r_{2\perp}}{\cos(\alpha+\theta')}\right)\\
  \label{eq:tautri}
  &\times
  \tau^\text{at}_{3}\left(\frac{m_3}M\vecz P';
    \frac{r_{3\perp}}{\cos(\alpha+\theta')}\right),
\end{align}
incorporates the complicated internal configuration of the trimer
molecule through the bound state wave functions. In particular, the
notation
\begin{subequations}
\label{eq:rj_perp}
\begin{eqnarray}
  r_{1\perp}&=&S_2\cos(\alpha+\theta')+\frac{m_2+m_3}M \rho_\perp\\
  r_{2\perp}&=&S_2\cos(\alpha+\theta')-\frac{m_1}M\rho_\perp+
  \frac{m_3}{m_2+m_3}r_\perp\\
  r_{3\perp}&=&S_2\cos(\alpha+\theta')-\frac{m_1}M\rho_\perp-
  \frac{m_2}{m_2+m_3}r_\perp
\end{eqnarray}
\end{subequations}
has been chosen to emphasize the geometrical meaning of the position
arguments of the atom transmission functions: The quantities
$r_{i\perp}/\cos(\alpha+\theta')$ can be interpreted as the positions
of the individual atoms projected onto the slit line $\mathcal S$
while the integration variable $S_2$ represents the projected center
of mass position $R_\perp/\cos(\alpha+\theta')$. The trimer
transmission function (\ref{eq:tautri}) is, therefore, simply the
product of the three atomic transmission functions averaged over the
wave functions of the incident and the outgoing bound trimer state.
This intuitive result is a straightforward extension of the case of
dimer diffraction \cite{SK_JPB35}.

\subsection{The trimer diffraction pattern of the deep grating}

Thanks to the formal coincidence of Eqs.~(\ref{eq:trimer_transampl3})
and (\ref{eq:trimer_slitfunction}) with their counterparts
Eqs.~(\ref{eq:atom_transampl_gra2}) and (\ref{eq:atom_slitfunction})
of atom diffraction the derivation of the $n$-th order relative
diffraction intensity $I_n^{\gamma\gamma'}$ for the incident bound
state $\phi_{\gamma'}$ and the outgoing bound state $\phi_{\gamma}$
can be carried over immediately.  Therefore, we write the trimer
diffraction intensities in the form
\begin{equation}
  \label{eq:trimer_intensity}
  I_n^{\gamma\gamma'}=I_0^{\gamma\gamma'}
  \left(\frac{P_{s_1}+P'_{s_1}}{2 P'_{s_1}}\right)^2
  \frac{| a^\text{tri}_{{\gamma\gamma'}}
    \left(\vecz P'; \Delta P_{s_2}\right)|^2}
  {| a^\text{tri}_{{\gamma\gamma'}}\left(\vecz P'; 0\right) |^2}.
\end{equation}
Contrary to the atom case, however, the trimer slit function depends
on the spatially extended trimer bound states $\phi_{\gamma}$ and
$\phi_{\gamma'}$ and, therefore, in general
$|a^\text{tri}_{{\gamma\gamma'}}\left(\vecz P'; 0\right) |^2<S_0^2$.

Eq.~(\ref{eq:trimer_intensity}) determines the diffraction intensities
of both elastic ($\phi_{\gamma}=\phi_{\gamma'}$) and inelastic
($\phi_{\gamma}\neq\phi_{\gamma'}$) processes. Earlier works on the
helium trimer \cite{HK_PRL84} as well as on van der Waals dimers
\cite{SK_JPB35} have shown that diffraction orders corresponding to
inelastic processes are typically less intense by five to six orders
of magnitude than those of elastic processes.  They are, therefore,
experimentally less relevant.  In the following we focus on elastic
processes.  Analogous to the procedure in Section
\ref{sec:atomtiltedgrating}, $I_n^{\gamma\gamma}$ can be approximated
by a two-term cumulant expansion. The cumulants $R^\pm_{\gamma,j}$ now
depend on the trimer state $\phi_{\gamma}$.  Moreover, because of the
threefold van der Waals interaction an additional term
$\Omega_\gamma=\frac12\text{\,Im\,}(R^+_{2,\gamma}-R^-_{2,\gamma})$
should be retained in the expansion for sufficient numerical accuracy.
Taking these generalizations into account the $n$-th order relative
intensity becomes
\begin{align}
  \nonumber
  &\frac{I_n}{I_0}=\left[\frac{P_{s_1}+P'_{s_1}}
    {2 P'_{s_1}   \sqrt{S_{\text{eff},\gamma}^2+\Delta_\gamma^2} }\right]^2
  \exp\left[-(\Delta P_{s_2}\Sigma_\gamma)^2/\hbar^2
    -\Delta P_{s_2} \Gamma_\gamma/\hbar\right]\\
  \label{eq:trimer_intensity_cumul}
  &\times\
  \frac{\sin^2[\Delta P_{s_2}S_{\text{eff},\gamma}/2\hbar]+
    \sinh^2[(\Delta P_{s_2}\Delta_\gamma/\hbar+ 
    \Delta P_{s_2}^2\Omega_\gamma/\hbar^2)/2]}
  {\left[\Delta P_{s_2}/2\hbar\right]^2}
\end{align}
where, analogous to Eqs.~(\ref{eq:momentum_angles}) and
(\ref{eq:momtransfk}), the momentum components are to be evaluated at
the incident angle $\theta'$ and the diffraction angle $\theta_n$ as
\begin{displaymath}
  P_{s_1}=|\vecz P'|\cos(\alpha+\theta_n),
  \quad
  P'_{s_1}=|\vecz P'|\cos(\alpha+\theta'), 
\end{displaymath}
and the momentum transfer parallel to the $s_2$-axis is given by
\begin{displaymath}
  \Delta P_{s_2}=|\vecz P'|\left[\sin(\alpha+\theta_n)-
    \sin(\alpha+\theta')\right].  
\end{displaymath}
Analogous to the atom case the effective slit width
$S_{\text{eff},\gamma}$ is related to the trimer transmission function
(\ref{eq:tautri}) by the equation
\begin{equation}
  \label{eq:Seff_trimer_1}
  S_{\text{eff},\gamma}=S_0-\text{\,Re\,} \int_{-{S_0}/2}^{{S_0}/2} d S_2
  \left[1-\tau^\text{tri}_{{\gamma\gamma}}(\vecz P'; S_2)
  \right].
\end{equation}

\begin{figure}[htbp]
  \centering
  \epsfig{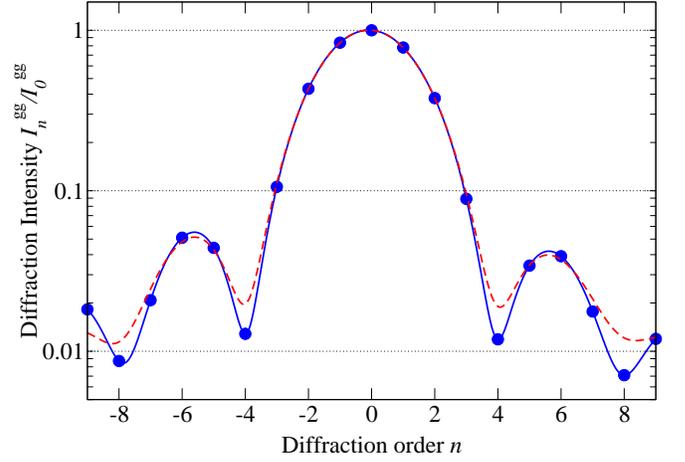}
  \caption{Diffraction intensities of a
    pure beam of ground state {\Hetrimer} at $v=500$\,m/s from a
    $d=100$\,nm transmission grating at $\theta'=21^\circ$ angle of
    incidence for the grating geometry of Fig.~\ref{fig:geometry_bar}.
    The solid curve was calculated using
    Eq.~(\ref{eq:trimer_intensity}), the dashed curve shows the
    two-term cumulant approximation (\ref{eq:trimer_intensity_cumul}).
    To guide the eye these functions are shown continuously.  The
    circles on top of the solid curve at integer $n$ mark the
    experimentally accessible diffraction orders
    $I^\text{gg}_n/I^\text{gg}_0$.}
  \label{fig:In_Trimer_Cumulants}
\end{figure}

Fig.~\ref{fig:In_Trimer_Cumulants} shows elastic diffraction
\begin{figure}[htbp]
  \centering
  \epsfig{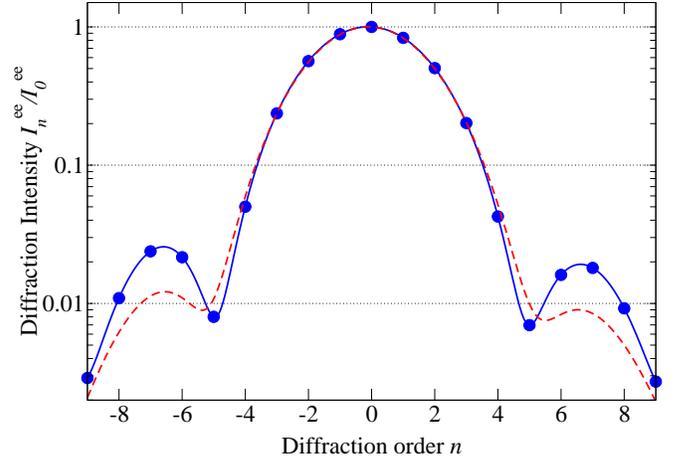}
  \caption{As in Fig.~\ref{fig:In_Trimer_Cumulants} but
    for a pure beam of excited state {\Hetrimer}. Due to the larger
    pair distance of the excited state the effective slit width is
    smaller, resulting in a considerably wider envelope function than
    for the ground state.}
  \label{fig:In_Trimer_Cumulants-Ef}
\end{figure}
intensities for a beam of {\Hetrimer} in its ground state calculated
according to Eqs.~(\ref{eq:trimer_intensity}) and
(\ref{eq:trimer_intensity_cumul}). The asymmetry of this diffraction
pattern is not as pronounced as in the atom case
(Fig.~\ref{fig:In_Atom_Cumulants}). This is due to the threefold mass
of the trimer which entails a three times shorter de Broglie
wavelength. Similarly, Fig.~\ref{fig:In_Trimer_Cumulants-Ef} shows
diffraction intensities for a beam of {\Hetrimer} in its excited
state.  

Since inelastic diffraction processes are negligible an experimental
diffraction pattern of a {\Hetrimer} beam will in general be well
described by an incoherent superposition of the individual diffraction
patterns of the two bound states weighted by their relative population
numbers in the beam.  In the following section we first derive the
trimer size effect for a pure beam containing trimers in only one state.
Hereafter the treatment of a mixed beam will be considered.

\section{How to determine the trimer size\label{sec:trimersize}}

\subsection{The trimer size effect}

Since the effective slit width of the trimer (\ref{eq:Seff_trimer_1})
depends, on the one hand, on the trimer bound state (through
$\tau^\text{tri}_{{\gamma\gamma}}(\vecz P'; S_2)$) and is, on the
other hand, experimentally accessible (through $I_n^{\gamma\gamma}$),
it represents a link between experiment and theory. Earlier work on
atom and dimer diffraction revealed that the difference
$S_0-S_{\text{eff},\gamma}$ carries information about both the van der
Waals surface interaction \cite{GSTHK_PRL83} and the size of weakly
bound dimers \cite{GSTHKS_PRL85}. The reduction of the slit width by
the dimer size was found to be $\frac12\expect{r}$ where $\expect{r}$
denotes the dimer bond length.  The subsequent evaluation of helium
dimer diffraction data yielded the experimental result
$\expect{r}=5.2\pm 0.4$\,nm for {\Hedimer} \cite{GSTHKS_PRL85}.

\begin{figure}[htbp]
  \centering
  \epsfig{file=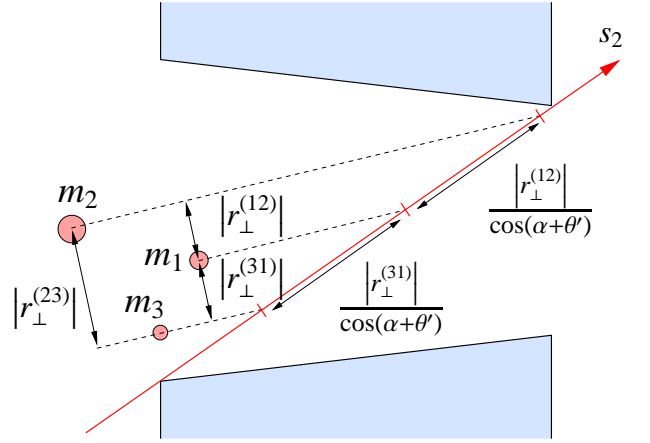,
    width=\columnwidth}
  \caption{Geometrical interpretation of the effective
    slit width formula. Clearly, the expression
    $\frac12(|r^{(23)}_\perp| + |r^{(31)}_\perp| +|r^{(12)}_\perp|)$
    may be interpreted as the ``width'' (projected diameter) of the
    trimer perpendicular to its incident direction.  Taking the
    expectation values with the bound state wave function yields the
    expression (\ref{eq:trimer_width}).  The multiplication by
    $[\cos(\alpha+\theta')]^{-1}$ corresponds to an orthogonal
    projection onto the slit line $\mathcal S$. Hence the slit width
    $S_0$ appears reduced by the projected width of the trimer.}
  \label{fig:Seff3-interpretation}
\end{figure}

In the following we derive the corresponding size effect for a trimer.
To this end we explicitly insert the trimer transmission function
(\ref{eq:tautri}) into Eq.~(\ref{eq:Seff_trimer_1}). By definition the
atom transmission functions $\tau^\text{at}_{i}(\vecz p'_i; s_{i2})$ in
Eq.~(\ref{eq:tautri}) are zero if their positional arguments
$s_{i2}=r_{i\perp}/\cos(\alpha+\theta')$ lie outside the slit interval
$[-S_0/2,S_0/2]$. This fact may be utilized to reduce the integration
interval for the center of mass position $S_2$: at fixed relative
coordinates $\vec\rho,\vec r$ the interval may be limited, for
$r_\perp>0$, to
\begin{subequations}
\label{eq:S_2_reduced_interval}
\begin{equation}
  \label{eq:S_2_r_perp_gt_0}
  -\frac{S_0}2+\frac{\Delta^+_2}{\cos(\alpha+\theta')} < S_2 <
  \frac{S_0}2-\frac{\Delta^+_1}{\cos(\alpha+\theta')}
\end{equation}
and, similarly, for $r_\perp < 0$, to
\begin{equation}
  \label{eq:S_2_r_perp_lt_0}
  -\frac{S_0}2+\frac{\Delta^-_1}{\cos(\alpha+\theta')} < S_2 <
  \frac{S_0}2-\frac{\Delta^-_2}{\cos(\alpha+\theta')}.
\end{equation}
\end{subequations}
Here, the geometrical quantities
\begin{align}
  \nonumber
  \Delta^\pm_1&=\pm\frac12
  \left\{
    \frac{m_2+m_3-m_1}M\rho_\perp+\frac{m_3}{m_2+m_3}r_\perp
  \right.\\
  \label{eq:Delta1pm}
  & \hspace{1cm}
  \left.
    \pm\left|\rho_\perp-\frac{m_3}{m_2+m_3}r_\perp\right|
  \right\}
\end{align}
and
\begin{align}
  \nonumber
  \Delta^\pm_2&=\pm\frac12
  \left\{
    -\frac{m_2+m_3-m_1}M\rho_\perp+\frac{m_2}{m_2+m_3}r_\perp
  \right.\\
  \label{eq:Delta2pm}
  &\hspace{1cm}
  \left.
    \pm\left|\rho_\perp+\frac{m_2}{m_2+m_3}r_\perp\right|
  \right\}
\end{align}
have been introduced. Neglecting, for the moment, the van der Waals
interaction, all atom transmission functions are unity inside the
reduced domain of integration (\ref{eq:S_2_reduced_interval}). In this
case the effective slit width depends only on $\Delta^\pm_{1}$ and
$\Delta^\pm_{2}$.  Accordingly, we call it the geometrical part of the
effective slit width and denote it by
\begin{align}
  \nonumber
  S^{\text{geom}}_{\text{eff},\gamma}=&S_0-\frac{1}{\cos(\alpha+\theta')}
  \text{\,Re\,}\int d^3\rho d^3 r\
  |\phi_{\gamma}(\vec\rho,\vec r)|^2
  \\
  \label{eq:Seffggeom_trimer}
  &\times
  \left\{
    \left[\Delta^+_1+\Delta^+_2\right]\Theta(r_\perp)
    +
    \left[\Delta^-_1+\Delta^-_2\right]\Theta(-r_\perp)\right\}
\end{align}
where $\Theta(r_\perp)$ is the Heaviside step function.  Both
integrands in Eq.~(\ref{eq:Seffggeom_trimer}) can be simplified using
the transformation properties of the Jacobi coordinates
(\ref{eq:jacobi-coordinates-pos-transf}) and the wave functions
(\ref{eq:wavefunctions_ijk_jki}). Combining the results, the
geometrical part of the effective slit width becomes
\begin{equation}
  \label{eq:Seffggeom_trimer2}
  S^{\text{geom}}_{\text{eff},\gamma}=S_0-
  \frac{
      \expect{\left|r^{(23)}_\perp\right|}_\gamma
      +\expect{\left|r^{(31)}_\perp\right|}_\gamma
      +\expect{\left|r^{(12)}_\perp\right|}_\gamma
  }{2\cos(\alpha+\theta')}
\end{equation}
where the expectation values in the numerator are defined as 
\begin{equation}
  \label{eq:rjk_erwartungswerte}
  \expect{\left|r^{(jk)}_\perp\right|}_\gamma=
  \int d^3\rho^{(i)} d^3 r^{(jk)}
  \left|\phi_{\gamma}^{(i,jk)}(\vec\rho^{(i)},\vec r^{(jk)})\right|^2
  \left|r^{(jk)}_\perp\right|.
\end{equation}
In Eq.~(\ref{eq:Seffggeom_trimer2}) the symmetric term
\begin{align}
  \label{eq:trimer_width}
    \frac{1}{2}
    \left(
      \expect{\left|r^{(23)}_\perp\right|}_\gamma
      +\expect{\left|r^{(31)}_\perp\right|}_\gamma
      +\expect{\left|r^{(12)}_\perp\right|}_\gamma
    \right)
\end{align}
represents the expectation value of the ``width'' (projected diameter)
of the trimer perpendicular to the incident direction. Therefore, as
the factor $[\cos(\alpha+\theta')]^{-1}$ corresponds to an orthogonal
projection of the perpendicular coordinates onto the slit line
(cf.~Fig.~\ref{fig:Seff3-interpretation})
$S^{\text{geom}}_{\text{eff},\gamma}$ is simply smaller than $S_0$ by
the projected width of the trimer.

In the presence of the van der Waals interaction an additional term
$S^{\text{vdW}}_{\text{eff},\gamma}$ accounting for the deviation from
unity of the atom transmission functions arises. The entire effective
slit width is the sum
\begin{equation}
  \label{eq:SeffggeomvdW}
  S_{\text{eff},\gamma}=S^{\text{geom}}_{\text{eff},\gamma}
  +S^{\text{vdW}}_{\text{eff},\gamma}.
\end{equation}
As the general expression for the van der Waals part
$S^{\text{vdW}}_{\text{eff},\gamma}$ is long and little informative we
shall not give its general form explicitly.

\subsection{The size effect for three identical Bosons}

\begin{figure}[htbp]
  \centering
  \epsfig{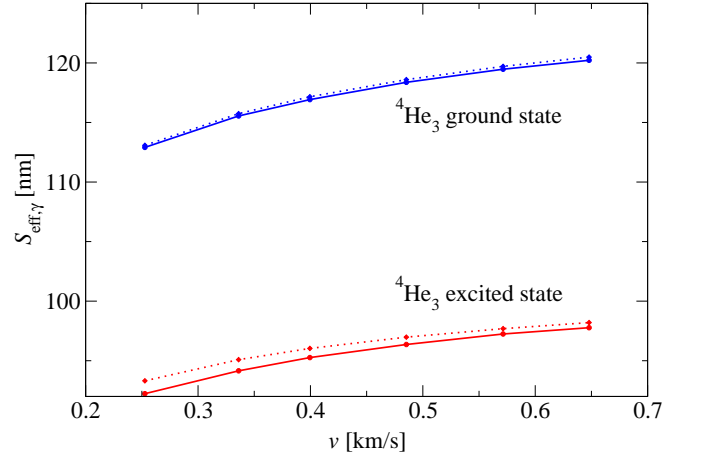}
  \caption{Effective slit widths versus the beam
    velocity $v=|\vec P'|/M$ for the ground state and the excited
    state of {\Hetrimer}. The numerical results using the full
    expression Eq.~(\ref{eq:SeffgvdW_trimer-suT}) are shown as solid
    curves and the approximation Eq.~(\ref{eq:SeffgvdW_trimer-suT-1})
    as dotted curves.  The angle of incidence was taken as
    $\theta'=21^\circ$.  The approximation becomes more reliable at
    high velocities as the impact of the van der Waals interaction
    decreases. Asymptotically, for high velocities, both pairs of
    curves approach their respective upper limits
    $S^{\text{geom}}_{\text{eff},\gamma}$ given by
    Eq.~(\ref{eq:Seffggeom_trimer2_suT_pre}).}
  \label{fig:Seff3aprx}
\end{figure}

In the remaining paragraphs of this section we focus on trimers of
three identical Bosons, such as the {\Hetrimer}. By consequence, we
denote the masses by $m=m_i$ and the equal projected pair distances by
$\expect{|r_\perp|}_\gamma$.  The geometric part of the effective slit
width (\ref{eq:Seffggeom_trimer2}) immediately reduces to
\begin{equation}
  \label{eq:Seffggeom_trimer2_suT_pre}
  S^{\text{geom}}_{\text{eff},\gamma}=S_0-\frac32
  \frac{\expect{\left|r_\perp\right|}_\gamma}{\cos(\alpha+\theta')}.
\end{equation}
Moreover, if the spatial extent of the bound state wave function is
small compared to the slit width, the van der Waals part
$S^{\text{vdW}}_{\text{eff},\gamma}$ is to very good approximation
given by
\begin{align}
  \nonumber
  &S^{\text{vdW}}_{\text{eff},\gamma}\simeq-\text{\,Re\,}
  \int d^3\rho d^3 r\
  |\phi_{\gamma}(\vec\rho,\vec r)|^2\\
  \nonumber
  &\left\{
    \int_0^{{S_0}/2}  d S'_2
    \left[1-
      \tau^\text{at}\left(\frac{{\vecz P}'}{3}; S'_2\right)
      \tau^\text{at}\left(\frac{{\vecz P}'}{3}; 
        S'_2-\frac{\left|\rho_\perp-\frac12|r_\perp|\right|}
        {\cos(\alpha+\theta')}\right)
    \right.\right.\\
  \nonumber
  & \left.\left.\ \ \ \times\ 
      \tau^\text{at}\left(\frac{{\vecz P}'}{3}; 
        S'_2-\frac{|r_\perp| + \left(\rho_\perp-\frac12|r_\perp|\right)
          \Theta\left(\rho_\perp-\frac12|r_\perp|\right)}
        {\cos(\alpha+\theta')}\right)
    \right]\right.\\
  \nonumber
  &+\!\!
  \int_{-{S_0}/2}^0  d S'_2
  \left.\left[1-
      \tau^\text{at}\left(\frac{{\vecz P}'}{3}; S'_2\right)
      \tau^\text{at}\left(\frac{{\vecz P}'}{3}; 
        S'_2+\frac{\left|\rho_\perp+\frac12|r_\perp|\right|}
        {\cos(\alpha+\theta')}\right)
    \right.\right.\\
  \label{eq:SeffgvdW_trimer-suT}
  & \left.\left.\ \ \ \times\
      \tau^\text{at}\left(\frac{{\vecz P}'}{3}; 
        S'_2+\frac{|r_\perp| - \left(\rho_\perp+\frac12|r_\perp|\right)
          \Theta\left(-\rho_\perp-\frac12|r_\perp|\right)}
        {\cos(\alpha+\theta')}\right)
    \right]\right\}.
\end{align}
Within the approximation (\ref{eq:SeffgvdW_trimer-suT}) it is evident
that $S^{\text{vdW}}_{\text{eff},\gamma}$ is indeed zero if the atom
transmission functions are unity inside the slit, and if the spatial
extent of the trimer wave function is small on the scale of the slit
width.  Therefore, if $S^{\text{vdW}}_{\text{eff},\gamma}$ were but a
small correction to the full effective slit width
(\ref{eq:SeffggeomvdW}) it could be neglected, and the projected
trimer pair distance $\expect{|r_\perp|}$ could be determined using
Eq.~(\ref{eq:Seffggeom_trimer2_suT_pre}).  Experience with dimer
diffraction has shown, however, that the effect of the van der Waals
interaction can be of the same order as the pair distance
\cite{GSTHKS_PRL85} and must be accounted for. Since
Eq.~(\ref{eq:SeffgvdW_trimer-suT}) depends on the full trimer wave
function it cannot be used immediately for the evaluation of
experimental data and an approximation is required.  The integrand in
Eq.~(\ref{eq:SeffgvdW_trimer-suT}) is, however, slowly varying on the
scale of the variation of $|\phi_{\gamma}(\vec\rho,\vec r)|^2$.
Therefore, the positional arguments of the atom transmission functions
can approximately be replaced by their expectation values. This
approach is in analogy to Ref.~\cite{GSTHKS_PRL85}. An analysis of the
combinations of $\rho_\perp$ and $r_\perp$, using once more the
transformation properties of the relative coordinates
(\ref{eq:jacobi-coordinates-pos-transf}), shows that these expectation
values are expressible solely in terms of $\expect{|r_\perp|}$. For
example,
\begin{align}
  \expect{\left|\rho_\perp-\frac12| r_\perp|\right|}_\gamma
  =\expect{\left| r_\perp\right|}_\gamma
\end{align}
and
\begin{align}
  \expect{\left(\rho_\perp-\frac12| r_\perp|\right)
      \Theta\left(\rho_\perp-\frac12| r_\perp|\right)}_\gamma
  =\frac14\expect{| r_\perp|}_\gamma.
\end{align}
Inserting these one finds as the final form of the van der Waals part
of the effective slit width
\begin{eqnarray}
  \nonumber
  \lefteqn{
    S^{\text{vdW}}_{\text{eff},\gamma}\simeq
    -\text{\,Re\,}\Bigg\{
    \int_0^{{S_0}/2} d S'_2 
    \left[1-
      \tau^\text{at}\left(\frac{{\vecz P}'}{3}; S'_2\right)
    \right.}& &\\
  \nonumber
  &\times&\!\left.
    \tau^\text{at}\left(\frac{{\vecz P}'}{3}; S'_2-
      \frac{\expect{|r_\perp|}_\gamma}{\cos(\alpha+\theta')}\right)
    \tau^\text{at}\left(\frac{{\vecz P}'}{3}; S'_2-
      \frac54\frac{\expect{|r_\perp|}_\gamma}{\cos(\alpha+\theta')}\right)
  \right]\\
  \nonumber
  &+&\!
  \int_{-{S_0}/2}^0 d S'_2 
  \left[1-
    \tau^\text{at}\left(\frac{{\vecz P}'}{3}; S'_2\right)
    \tau^\text{at}\left(\frac{{\vecz P}'}{3}; S'_2+
      \frac{\expect{|r_\perp|}_\gamma}{\cos(\alpha+\theta')}\right)
  \right.\\
  \label{eq:SeffgvdW_trimer-suT-1}
  &\times&\!\left.
    \tau^\text{at}\left(\frac{{\vecz P}'}{3}; S'_2+
      \frac54\frac{\expect{|r_\perp|}_\gamma}{\cos(\alpha+\theta')}\right)
  \right]\Bigg\}.
\end{eqnarray}
In order to test the validity of this approximation we carried out a
numerical analysis of the error introduced by the replacement of
Eq.~(\ref{eq:SeffgvdW_trimer-suT}) by
Eq.~(\ref{eq:SeffgvdW_trimer-suT-1}): if applied to experimental data
the approximation entails, for the two theoretically predicted bound
states of {\Hetrimer}, a systematic overestimation of
$\expect{|r_\perp|}$ by 7\% ({\Hetrimer} ground state), or 3\%
(excited state). As seen from Fig.~\ref{fig:Seff3aprx} the
approximation is more reliable at high velocities as the impact of the
van der Waals interaction becomes smaller.

Theoretical studies of the helium trimer commonly state the
expectation value of the pair distance $\expect{r}$ itself, where
$r=|\vec r|$, rather than a component such as $\expect{|r_\perp|}$.
To link experimental results to these, a relation between $\expect{r}$
and $\expect{|r_\perp|}$ must be established.  Both predicted
{\Hetrimer} bound states are spherically symmetric (zero total angular
momentum).  Moreover, the two-body scattering matrix corresponding to
the He-He potential is dominated by the shallow $s$-wave bound state
pole of {\Hedimer}, and higher partial waves may to good approximation
be neglected \cite{NL_PRA28}. By the Faddeev equations
(e.g.~Ref.~\cite{sitenko}) for the helium trimer bound state, it is
then possible to derive the relation
\begin{equation}
  \label{eq:rperp_r2}
  \expect{|r_\perp|}=\frac12\expect{r}.
\end{equation}
In summary, the effective slit width (\ref{eq:SeffggeomvdW}) depends
to good approximation only on one trimer parameter, namely the
expectation value of the pair distance $\expect{r}$. Consequently,
$\expect{r}$ can, in principle, be determined from trimer diffraction
data.

\subsection{Experimental considerations}

Using the results of the preceding sections the improvement in
resolution through diffraction at non-normal incidence over normal
incidence may be estimated. The evaluation process of experimental
data involves two main steps: Firstly, values for the effective slit
width $S_{\text{eff},\gamma}$ must be obtained by numerical fits of
the intensity formula (\ref{eq:trimer_intensity_cumul}) to
experimental diffraction patterns. Secondly, $\expect{r}$ is
determined from a fit of Eq.~(\ref{eq:SeffggeomvdW}) to the values for
$S_{\text{eff},\gamma}$.  The stronger the dependence of this
procedure on $\expect{r}$ the more precisely $\expect{r}$ can be
determined. As $\expect{r}$ changes the width of the Kirchhoff-like
slit function in Eq.~(\ref{eq:trimer_intensity_cumul}) a natural
measure for this dependence is provided by the number of diffraction
orders under the central maximum of the slit function to either side
of the forward direction.  This number, which we denote here by $n_c$,
and which we treat as a continuous variable, can be approximately
written as
\begin{equation}
  \label{eq:number_nc}
  n_c=\frac{d_\perp}{s_\perp}
\end{equation}
where $d_\perp$ denotes the (projected) period perpendicular to the
beam and $s_\perp$ denotes the projected slit width. At the angle of
incidence $\theta'$ the projected period is $d_\perp=d\cos\theta'$.
Furthermore, neglecting for this estimate the van der Waals part, we
insert
$s_\perp=S^{\text{geom}}_{\text{eff},\gamma}\cos(\alpha+\theta')$ for
the projected slit width of the deep grating where
$S^{\text{geom}}_{\text{eff},\gamma}$ will be taken from
Eq.~(\ref{eq:Seffggeom_trimer2_suT_pre}). The relative variation of
$n_c$ with $\expect{r}$ can then be calculated and becomes, to leading
order in $\expect{r}/S_0$,
\begin{displaymath}
  \frac{1}{n_c}{\frac{ \text{d} n_c}{ \text{d} \expect{r}}}
  \simeq \frac34\frac{1}{S_0\cos(\alpha+\theta')}.
\end{displaymath}
In contrast, at normal incidence the right hand side would be
$3/(4s_0)$. Inserting the parameters of
Fig.~\ref{fig:In_Trimer_Cumulants} yields $(\text{d}
n_c/\text{d}\expect{r})/n_c\approx 2.6\times10^{-2}\,\text{nm}^{-1}$
for $\theta'=21^\circ$ as compared to
$1.2\times10^{-2}\,\text{nm}^{-1}$ for normal incidence. This roughly
twofold gain in sensitivity is expected to halve the final error bars
on $\expect{r}$.

Finally, since the population ratio of the two predicted {\Hetrimer}
states in the nozzle beam is generally unknown the situation of a
mixed beam must be considered. To analyze this we have summed
diffraction patterns as shown in Figs.~\ref{fig:In_Trimer_Cumulants}
and \ref{fig:In_Trimer_Cumulants-Ef} for different population ratios.
Hereafter, we have used Eq.~(\ref{eq:trimer_intensity_cumul}) to
determine, from the summed patterns, an average effective slit width
$S_\text{eff}$ and from this an average bond length $\expect{r}$. It
turned out that the such determined $\expect{r}$ varies almost
linearly with the population ratio from the ground state value of
$\expect{r}$ (pure ground state beam) to the excited state value (pure
excited state beam). Therefore, three possible outcomes of an
experiment are to be expected.  A value of $\expect{r}\approx1$\,nm
would be attributed to the ground state and indicate a negligible (or
zero) population of the excited state.  Equivalently, a result of
about $8$\,nm would doubtlessly provide evidence for the excited state
and its large pair distance.  Thirdly, a value in between these two
would indicate that both states are present and evidence for the
excited state would still be available. A controlled variation of the
accessible beam parameters might then allow to influence the
population ratio in favor of either state and to measure the pair
distance for one state with less disturbance by the other.

\section{Conclusions\label{sec:summary}}

Motivated by the long-standing interest in the Efimov effect
\cite{E_PL33B} we have studied the diffraction of weakly bound trimers
in a typical matter optics setup. As an earlier diffraction experiment
for the spatially extended helium dimer ($\expect{r}=5.2$\,nm)
\cite{GSTHKS_PRL85} had indicated that the resolution provided by a
custom $s_0=60$\,nm transmission grating, at normal incidence, may be
insufficient to resolve the helium trimer ground state
($\expect{r}=0.96$\,nm predicted \cite{BK_PRA64}) it had suggested
itself to use oblique (non-normal) incidence at a rotated transmission
grating for reducing the projected slit width.  The partial shadowing
of the slits caused by the finite thickness of the etched material
grating has required, however, a revision of the theory of atom
diffraction.  In particular, the familiar mirror symmetry encountered
in diffraction patterns from normal incidence is lifted for non-normal
incidence. This effect was visible in Fig.~5 of
Ref.~\cite{GSTMSS_PRA61} but went unnoticed. It has been traced back
to the non-alignment of the direction of periodicity of the grating
with the shadow lines of its bars, or, equivalently, its slit lines.
The weak attractive van der Waals surface interaction, which
introduces an additional but minor asymmetry, has been taken into
account in a way similar to the case of normal incidence.

Using atom diffraction as one building block, the multi-channel
many-body quantum mechanical scattering theory approach of
Refs.~\cite{HK_PRA57,HK_PRA61} has been extended to derive the
constitutive formulas of trimer diffraction.  While this procedure
structurally partly parallels that of dimer diffraction it is
mathematically more complex due to the additional atom. The resulting
equations for the trimer diffraction pattern, however, have been
readily interpretable and provide intuitive physical insight into
diffraction of weakly bound molecules: the significant measurable
quantity is the quantum mechanical expectation value of the ``width''
(projected diameter) of the trimer perpendicular to its flight
direction. For identical Boson trimers, such as the helium trimer, the
width is related to the molecular bond length by $\frac34\expect{r}$.
This fact can be used, in principle, to determine $\expect{r}$ from a
matter diffraction experiment.

If a transmission grating of the type used earlier by Grisenti
\emph{et al} \cite{GSTHK_PRL83} is rotated by $\theta'=21^\circ$ the
projected slits appear approximately half as wide as the nominal slits
of the grating. This leads to an estimated doubling of the resolution,
sufficient to determine the ground state pair distance of {\Hetrimer}.
Moreover, should the {\Hetrimer} Efimov state, whose pair distance is
predicted to be larger by almost an order of magnitude
\cite{BK_PRA64}, exist and should its population in the beam be
significant, it ought to be clearly distinguishable from the ground
state solely by its size.

\begin{acknowledgments}
  We would like to thank R.\ Br{\"uhl}, O.\ Kornilov, A.\ Kalinin,
  T.~K{\"o}hler and J.P.~Toennies for stimulating discussions. This
  research was supported by the Deutsche Forschungsgemeinschaft.
\end{acknowledgments}

\appendix

\section{Surface interaction\label{sec:surface_interaction}}

Earlier work has shown that at beam velocities typically encountered
in matter diffraction experiments the effective reduction of the slit
width due to both the van der Waals surface interaction and the finite
molecular size can be of the same order of magnitude
\cite{GSTHK_PRL83, GSTHKS_PRL85}. A quantitative determination of the
atom transmission function $\tau^\text{at}(\vecz p'; s_2)$, which was
inserted into Eq.~(\ref{eq:atom_slitfunction}), is therefore
necessary. As in Ref.~\cite{GSTHK_PRL83} we use the eikonal
approximation to write $\tau^\text{at}(\vecz p'; s_2)=\exp[ i
\varphi(\vecz p'; s_2)]$ for $s_2$ inside the slit, and
$\tau^\text{at}(\vecz p'; s_2)=0$ outside.  The phase function
$\varphi(\vecz p'; s_2)$ is given by \cite{joachain}
\begin{equation}
  \label{eq:phase_function}
  \varphi(\vecz p'; s_2)=-(\hbar v)^{-1}\int d t\ W_\text{surf}(\vecz s(t))
  ,\quad v=\frac{|\vecz p'|}m,
\end{equation}
where the straight path of integration $\vecz s(t)$ must be taken to
run parallel to the direction of incidence, and to cross the slit line
$\mathcal S$ at the position $s_2$. The surface interaction
$W_\text{surf}(\vecz x)$ at a position $\vecz x$ between two bars is
calculated from the integration of an attractive $-C_6/l^6$ potential
of Lennard-Jones type (the repulsive part has already been modeled by
the boundary conditions in Sec.~\ref{sec:generaltheory}) over the
volume of the bars. Carrying out all four integrations for the typical
wedge-shaped bars shown in Fig.~\ref{fig:geometry_bar} the phase
function can be calculated explicitly. Using the abbreviations
\begin{displaymath}
  C_3=\frac{\pi C_6}6,\quad
  \widetilde{d}=\frac{\cos\theta'}{\cos(\alpha+\theta')}d,\quad
  \widetilde{s_0}=\frac{\cos\theta'}{\cos(\alpha+\theta')}s_0,
\end{displaymath}
it reads
\begin{eqnarray*}
  \nonumber
  \lefteqn{\varphi(\vecz p';s_2)=
    \frac{C_3}{2\hbar v\cos^2\theta'\cos^2(\alpha+\theta')}
    }& &\\
  \nonumber
  &\times& \left\{
    \frac{\xi_{11}^{-2}+(\xi_{11}-\widetilde{d})^{-2}
      -\xi_{12}^{-2}-(\xi_{12}+\widetilde{d})^{-2}}{\tan\theta'+\tan\beta}
  \right.\\
  \label{eq:phasenfunktion-2}
  & &
  \left. +
    \frac{\xi_{21}^{-2}+(\xi_{21}-\widetilde{d})^{-2}
      -\xi_{22}^{-2}-(\xi_{22}-\widetilde{d})^{-2}}{\tan\theta'-\tan\beta} 
  \right\}
\end{eqnarray*}
where $\xi_{11}=\frac{S_0}2-s_2$, $\xi_{21}=\frac{S_0}2+s_2$,
$\xi_{12}=\frac{S_0}2-s_2-\widetilde{s_0}+
\frac{\cos((\alpha+\theta')-2\theta')}{\cos(\alpha+\theta')}S_0$, and
$\xi_{22}=\frac{S_0}2-s_2-\widetilde{s_0}$.


\begin{thebibliography}{52}
\bibitem{footnote-HN} {Since $N\gg 1$ one may replace $H_{N-1}$ by
    $H_N$ in Eq.~(\ref{eq:atom_transampl_gra2}).}
\bibitem{footnote-othertrimers} {The theory of trimer diffraction
    derived in Sections \ref{sec:trimerscatteringtheory} and
    \ref{sec:trimertiltedgrating} also applies if one or more of the
    atoms are substituted by a more strongly bound molecule.  In the
    Ar$_2$--HCl cluster, for example, the binding between the hydrogen
    and the chlorine atom is comparatively tight \cite{KCREG_JCP86}.
    It is however required that electronic or rovibrational
    excitations of the molecule cannot occur at the energies
    available in a nozzle beam (a few tens of meV).}
\end{thebibliography}
\end{document}